\documentclass[preprint,12pt]{elsarticle}

\usepackage{natbib}
\usepackage{amsmath,amssymb,amsfonts}
\usepackage{algorithmic}
\usepackage{graphicx}
\usepackage{textcomp}
\usepackage{hyperref}
\usepackage{siunitx}

\usepackage{gensymb}
\usepackage{float}
\usepackage{enumitem} 
\usepackage{accents}
\newcommand{\via}{{\textit{via}~}}
\newcommand{\ie}{{\textit{i.e.}~}}
\newcommand{\apriori}{{\textit{a priori}~}}
\newcommand{\rmi}[1][]{\mathrm{i}}

\newcommand{\Real}{{\mathrm{Re}}}

\newcommand{\Imag}{{\mathrm{Im}}}

\newcommand{\sio}[1][]{$\mathrm{SiO}_2$~}
\newcommand{\tio}[1][]{$\mathrm{TiO}_2$~}
\newcommand{\ta}[1][]{$\mathrm{Ta}_2\mathrm{O}_5$}

\newcommand{\norm}[1]{\left \lVert #1 \right \rVert}

\usepackage[normalem]{ulem}
\usepackage{xcolor}
\newcommand{\revision}[1]{{\textcolor{black} {#1}}}

\DeclareMathOperator*{\argmin}{arg\,min}

\journal{}

\begin{document}

\begin{frontmatter}

\title{SEMPO - Retrieving complex poles, residues and zeros from arbitrary real spectral responses\tnoteref{t1}}

\tnotetext[t1]{This work was funded by the French National Research Agency ANR
Project DILEMMA (ANR-20-CE09-0027)}

\author[1]{Isam BEN SOLTANE\corref{cor1}}
\ead{isam.ben-soltane@fresnel.fr}
\author[1]{Mahé ROY}
\author[1]{Rémi ANDRE}
\author[1]{Nicolas BONOD\corref{cor1}}
\ead{nicolas.bonod@fresnel.fr}

\affiliation[1]{Aix Marseille Univ, CNRS, Centrale Mediterranee, Institut Fresnel, 13013 Marseille, France}

\cortext[cor1]{Corresponding authors}

\begin{abstract}
The Singularity Expansion Method Parameter Optimizer - SEMPO - is a toolbox to extract the complex poles, zeros and residues of an arbitrary response function acquired along the real frequency axis. SEMPO allows to determine this full set of complex parameters of linear physical systems from their spectral responses only, without prior information about the system. The method leverages on the Singularity Expansion Method of the physical signal. This analytical expansion of the meromorphic function in the complex frequency plane motivates the use of \revision{an accuracy-driven improved version of the Cauchy method constrained by properties of physical systems, as well as an auto-differentiation-based optimization approach}. Both approaches can be sequentially associated to provide highly accurate reconstructions of physical signals in large spectral windows. The performances of SEMPO are assessed and analysed in several configurations that include the dielectric permittivity of materials and the optical response spectra of various optical metasurfaces. \revision{SEMPO's performances are thoroughly analyzed and benchmarked with other state-of-the-art methods to highlight its capability to retrieve the natural poles of a physical system}. 

\section*{Program summary}

\begin{itemize}
    \item Program title: Singularity Expansion Method Parameter Optimizer (SEMPO)
    \item CPC Library link to program files:
    \item Developer's repository link: https://doi.org/10.5281/zenodo.15210008
    \item Licensing provisions: Creative Commons Attribution 4.0 International
    \item Programming language: Python
    \item Nature of problem: Spectral functions of scattering coefficients of physical systems can be rigorously expanded through the Singularity Expansion Method. This expansion method permits to cast the spectral function through two equivalent forms: a sum of complex rational functions depending on complex poles and residues, and a factorized expression that depends on complex poles and zeros. Spectral functions are usually acquired along the real frequency axis, the challenge is thus to identify the set of complex poles, zeros and residues from these acquisitions at real frequencies.
    \item Solution method: SEMPO is a numerical toolbox that aims at extracting the complex poles, zeros and residues from arbitrary spectral functions. SEMPO relies on both an improved algebraic Cauchy method associated with the factorized expression, and an open-source auto-differentiation method associated with the expansion in terms of poles and residues. Both methods are based on the Singularity expansion method and leverage the properties of physical systems and signals. The accuracy of SEMPO is thoroughly assessed and analysed in different configurations including the spectral responses of optical metasurfaces and the dielectric permittivity of optical materials. 
\end{itemize}

\end{abstract}

%% Keywords
\begin{keyword}
Auto-differentiation\sep Cauchy method\sep complex analysis\sep poles\sep residues\sep singularity expansion method\sep zeros
\end{keyword}

\end{frontmatter}

\section{Introduction}

\noindent
The spectral response of a linear physical system carries rich information on the system and on its interaction with excitation fields.  The complex frequency space is well suited to perform this analysis, and the poles \revision{and zeros} of the different transfer operators or scattering channels provide much of this information on the system. Different methods have been developed in order to model the response of the system to a given excitation with parameters determined in the complex frequency space including the Singularity Expansion Method (SEM), the Resonant State Expansion, or the quasi-normal modes theory~\cite{baum_singularity_1971,grigoriev_optimization_2013,grigoriev_singular_2014,colom_modal_2018,ben_soltane_multiple-order_2023,sauvan2013theory,vial2014quasimodal,doost2014resonant,lalanne2019quasinormal,stout2021spectral,sauvan2022normalization}. \revision{The development of these methods has been motivated by the need to tailor and optimize resonances that are cumbersome in wave physics; Resonances are at the core of cavity-driven systems in water waves~\cite{meylan2012complex,bennetts2021complex,euve2023perfect}, acoustics~\cite{zalalutdinov2021acoustic}, or cavity opto-mechanics~\cite{barzanjeh2022optomechanics,vijayan2024cavity}. They play a key role in optics and photonics in enhancing light-matter interactions~\cite{lalanne2018light,bidault2019dielectric,bonod2020all,zerulla2022multi,do2024room,biechteler2025fabrication}, cavity electrodynamics~\cite{haroche2020cavity}, or designing highly efficient laser cavities~\cite{contractor2022scalable}. The strength of resonances is characterized by the quality factor that can be calculated through several methods. It can be first estimated under the assumption of well-defined Lorentzian spectral responses by taking the ratio between the central resonant frequency and the
full width at half maximum of the resonant response~\cite{zambrana2024quality}. It can also be calculated with the temporal coupled mode theory through the calculation of the radiative
and internal loss times~\cite{fan2003temporal,berguiga2021ultimate} or in the harmonic regime by calculating the ratio between the real and the imaginary part of a complex singularity~\cite{neviere1980homogeneous,wu2021nanoscale}. Recently, it was showed that the concept of quality factor can be generalized by a meromorphic quality function proportional to the Wigner-Smith time delay~\cite{soltane2024extracting}. An SEM expansion of this function shows that both poles and zeros must be taken into account to characterize resonances. This function identifies with the quality factor at frequencies equal to the real frequency of the singularity. These results illustrate the importance of retrieving poles, zeros and residues in the complex frequency plane.\\}

\noindent
The SEM provides an expansion of any physical signal, such as the elements of a scattering operator, in terms of its poles and residues or alternatively in terms of its poles and zeros determined in the complex frequency plane. %The problem of the retrieval of the poles, zeros and residues of a signal has been tackled by several groups over the past few years, for the different prospects it offers. 
The expressions resulting from the distributions of \revision{poles, zeros and residues} can be used to approximate functions at any frequency over large spectral windows, with small and discrete sets of parameters~\cite{desoer_zeros_1974,pond_determination_1982,kottapalli_accurate_1991,adve_application_1997,betz_efficient_2024,betz2025uncovering,binkowski2025resonancemodes}. They provide simple models to describe the behaviour of scatterers and cavities~\cite{sarkar_application_2000,colom_modal_2018,zaky_comparison_2020,ben_soltane_derivation_2022}. The behaviour of physical systems and their interaction with excitation signals can be explained in terms of poles and zeros~\cite{garcia-vergara_extracting_2017,chen_use_2022,ferise_exceptional_2022,colom_crossing_2023,ferise_optimal_2023,ben_soltane_generalized_2024}.\\

\noindent
\revision{If the complex frequency analysis turns out to be highly relevant for analyzing physical spectra and for characterizing interactions of waves with physical systems, data are in most cases experimentally acquired along the real frequency axis. The challenge is therefore to extract complex parameters from signals acquired along real frequencies. The SEM shows that any linear physical spectral response can be expanded onto an infinite set of complex poles, zeros and residues.}
\revision{It turns out that the distribution of poles, zeros and residues in the complex frequency plane determines the spectral shape of the response function over the real frequencies. This means that the information about their complex distribution is encoded in the spectral response on the real axis, and the challenge is therefore to extract this information from functions of real frequencies.} Different methods have been proposed to retrieve the \revision{poles, zeros and residues} from a spectral response either obtained in the complex $\omega$ plane or along the real frequency axis. When the available data are acquired or generated in the complex frequency plane, the calculation of the poles and residues (or equivalently the poles and zeros) is quite straightforward through contour integrals~\cite{beyn_integral_2012,van_barel_nonlinear_2016,colom_crossing_2023,binkowski_poles_2024}. However, in practice, many spectra are acquired over a real spectral range $[\omega_A, \omega_B]\subset\mathbb{R}$. In the latter case, much fewer methods exist but this longstanding problem has been attracting more and more attention~\cite{hua_matrix_1988,mandelshtam_harmonic_1997,gustavsen_rational_1999,lee_computation_2012,meylan2012complex,valera-rivera_aaa_2021,betz2025uncovering,binkowski2025resonancemodes,hofreither2021algorithm}. Existing methods mainly approximate the response functions as meromorphic functions, the parameters of which are either the poles and zeros (Cauchy method~\cite{lee_computation_2012} and AAA-\revision{Adaptive Antoulas-Anderson method-}~\cite{valera-rivera_aaa_2021} indirectly), or the poles and residues (matrix pencil method~\cite{hua_matrix_1988}, harmonic inversion~\cite{mandelshtam_harmonic_1997}, \revision{phase analysis~\cite{kowalczyk2018global}}, vector fitting~\cite{gustavsen_rational_1999}). They rely on matrix algebra, but can be quite sensitive to an initial guess of some parameters, as in the case of the vector fitting method.\\

\noindent
Here, we describe two approaches to identify, in the complex frequency plane, the \revision{poles, zeros and residues} of a physical system from spectra acquired over a finite range of real frequencies. The first approach is based on the Cauchy method~\cite{adve_application_1997,lee_computation_2012,sarkar_cauchy_2021} and works by expressing the function through a rational approximation. In this work, we describe the working principle of the original method and explain how it can be improved and optimized \revision{as to first minimize the error with a target signal sample in Section 2.1, and then to leverage the properties of physical systems to yield physics-informed corrections in Section 2.2.} The second method relies on auto-differentiation~\cite{paszke_automatic_2017,minkov_inverse_2020,so_revisiting_2022,alagappan_group_2023,so_multicolor_2023,ben_soltane_generalized_2024} to solve optimization problems through gradient-descent. The formulation of these optimization problems is detailed, as well as their application to retrieve the SEM. 
\revision{We apply and analyze the performances of these methods by studying the specular optical response of metasurfaces consisting of periodic nano-elements composed of dispersive materials. We consider 2D arrays of silicon nanodisks on a glass substrate and silver square nanopillars on a silver substrate, and a 1D gold grating consisting of nanoslits etched on a gold substrate. All photonic nanostructures exhibit rich optical spectra useful to assess the performances of SEMPO. Optical responses of these periodic photonic nanostructures are calculated with the Rigorous Coupled Wave Analysis implemented with the software Planopsim and the simulated data are provided in the repository link. Another category of optical response is considered by studying the dielectric permittivity of gold.}  \\

\noindent
The physical signal that we consider is a transfer function such as an element of the scattering $S$-matrix~\cite{demesy2018scattering,beutel2024treams}. Each element of the $S$-matrix is represented by a meromorphic function $h(\omega)$ which can be expanded using the SEM~\cite{baum_singularity_1971,grigoriev_exact_2011,colom_modal_2018,colom_modal_2019,ben_soltane_derivation_2022,ben_soltane_multiple-order_2023}, \ie
\begin{equation}
    h(\omega) = h_{NR} + \sum_\ell \frac{r^{(\ell)}}{\omega - p^{(\ell)}.}
\label{eq:5_1_sem}
\end{equation}
$h_{NR}$ is the non-resonant term, $\{p^{(\ell)}\}_\ell$ is the set of simple poles of $h(\omega)$ (we assume that the singularities are all poles of order $1$), and $r^{(\ell)}$ is the residue associated with the pole $p^{(\ell)}$. Equivalently, $h(\omega)$ can be expressed through the Singularity and Zero Factorization (SZF)~\cite{grigoriev_optimization_2013,grigoriev_singular_2013,grigoriev_singular_2014,ben_soltane_multiple-order_2023}:
\begin{equation}
    h(\omega) = \eta_0 ~ \frac{\displaystyle \prod_{\ell}(\omega - z^{(\ell)})}{\displaystyle \prod_{\ell}(\omega - p^{(\ell)})} 
\label{eq:5_2_szf}
\end{equation}
with $z^{(\ell)}$ the zeros of $h(\omega)$, and $\eta_0$ a known constant which depends on $z^{(\ell)}$, $p^{(\ell)}$, and the response $h(a)$ to an arbitrary frequency $a$. \revision{Let us point out that in lossless stable systems, \ie when the amplitude of $h(\omega)$ is constant over the real frequency axis, as is the case as a first approximation with the reflection coefficient of a perfect mirror for instance, then the poles and zeros are complex conjugates. In this case, the SZF identifies with the Blaschke product~\cite{meylan2017extraordinary}.}\\

\noindent
The distribution of poles $p^{(\ell)}$, zeros $z^{(\ell)}$, and residues $r^{(\ell)}$ fully characterize the response function $h(\omega)$.
Therefore, the \revision{poles and zeros} are intrinsic properties of physical systems, and their retrieval is relevant. However, there is a potentially infinite number of parameters, which implies that they are never all taken into account since the SZF and SEM must be numerically truncated. In the following we describe how to retrieve the poles, residues and zeros once we perform this truncature. Furthermore, we show how to combine both methods to improve the overall performance.

\section{The Cauchy method for rational approximations}

\subsection{Overview of the original method}

\noindent
Let us first present an overview of the Cauchy method~\cite{adve_application_1997,lee_computation_2012,sarkar_cauchy_2021} to retrieve the poles and zeros of a function from a set of measurements acquired along a real finite spectral range.\\ 

\noindent
We consider Equation~\eqref{eq:5_1_sem} truncated at the $M_p^{\mathrm{th}}$ order for the poles, or Equation~\eqref{eq:5_2_szf} also truncated at the $M_z^{\mathrm{th}}$ order for the zeros. Since the response function $h(\omega)$ is meromorphic, it can also be written as a rational function:
\begin{equation}
    h(\omega) = \frac{f(\omega)}{g(\omega)}
\label{eq:5_3}
\end{equation}
with $g(\omega)$ a polynomial of degree $M_p$, and $f(\omega)$ a polynomial of degree $M_z$. The degree $M_p$ of $g(\omega)$ is initialized to the number of expected poles, and is case dependent. The roots of $g(\omega)$ and $f(\omega)$ are respectively the poles and zeros of $h(\omega)$. The calculation of the poles and zeros is thus tantamount to the calculation of the coefficients $\mathbf{a}=(a_0, a_1, ..., a_{M_z})^T$ of $f(\omega)$ and $\mathbf{b}=(b_0, b_1, ..., b_{M_p})^T$ of $g(\omega)$. For any frequency $\omega$, we have
\begin{equation}
    f(\omega) = \mathbf{x}^{(z)}  \mathbf{a}
\label{eq:5_4}
\end{equation}
and
\begin{equation}
    g(\omega) = \mathbf{x}^{(p)}  \mathbf{b} 
\label{eq:5_5}
\end{equation}
with $\mathbf{x}^{(z)}=(1, \omega, \omega^2, ..., \omega^{M_z})$, $\mathbf{x}^{(p)}=(1, \omega, \omega^2, ..., \omega^{M_p})$. Equation~\eqref{eq:5_3} can thus be recast as
\begin{align}
    \left[ \mathbf{x}^{(z)}, ~ -h(\omega)\mathbf{x}^{(p)} \right]  \begin{bmatrix}
        \mathbf{a}\\
        \mathbf{b}\\ 
   \end{bmatrix} = 0
\label{eq:5_6}
\end{align}

\noindent
The data are acquired at $N$ frequencies $\omega_n$, $1\leq n\leq N$, between $\omega_{inf}$ and $\omega_{sup}\in\mathbb{R}$: 
\begin{equation}
    \omega_{inf} \leq \omega_1 < \omega_2 < ... < \omega_N \leq \omega_{sup}
\end{equation}
Let us write $h(\omega_n)=h_n$ the value of the response function at $\omega_n$. The following matrix notation of Equation~\eqref{eq:5_6} is obtained:
\begin{equation}
    [\mathbf{A}, -\mathbf{B}] \begin{bmatrix}
                                  \mathbf{a}\\
                                  \mathbf{b}\\
                              \end{bmatrix} = \mathbf{0}.
\label{eq:5_7}
\end{equation}
The matrix $\mathbf{A}$ reads as
\begin{equation}
    \mathbf{A} = \begin{bmatrix}
                      1 & \omega_1 & ... & \omega_1^{M_z}\\
                      1 & \omega_2 & ... & \omega_2^{M_z}\\
                      &...& \\
                      1 & \omega_N & ... & \omega_N^{M_z}\\
                 \end{bmatrix}\\
 \label{eq:5_8}
 \end{equation}
 
and $\mathbf{B}$ as

\begin{equation}
    \mathbf{B} = \begin{bmatrix}
                  h_1 & h_1\omega_1 & ... & h_1\omega_1^{M_p}\\
                  h_2 & h_2\omega_2 & ... & h_2\omega_2^{M_p}\\
                  &...& \\
                  h_N & h_N\omega_N & ... & h_N\omega_N^{M_p}\\
                 \end{bmatrix}.
\label{eq:5_9}
\end{equation}
\revision{While the polynomial coefficients $\mathbf{a}$ and $\mathbf{b}$ could be obtained by solving Equation~\eqref{eq:5_7} through the least-square method, the} Cauchy method instead constrains the kernel of the matrix $\mathbf{C}=[\mathbf{A},~ -\mathbf{B}]$ in order to obtain a unique algebraic solution up to a scaling factor. The number of unique frequencies $N$ is usually greater than the number of columns $2+M_p+M_z$ of $\mathbf{C}$. It follows that the rank $r$ of $\mathbf{C}$ is given by
\begin{equation}
    r = M_z+M_p+2 - K
\label{eq:5_10}
\end{equation}
with $K$ the dimension of its kernel. The Cauchy method aims at setting the dimension of the kernel to $K=1$.\\

\begin{figure}[ht!]
    \centering
    \includegraphics[width=.98\textwidth]{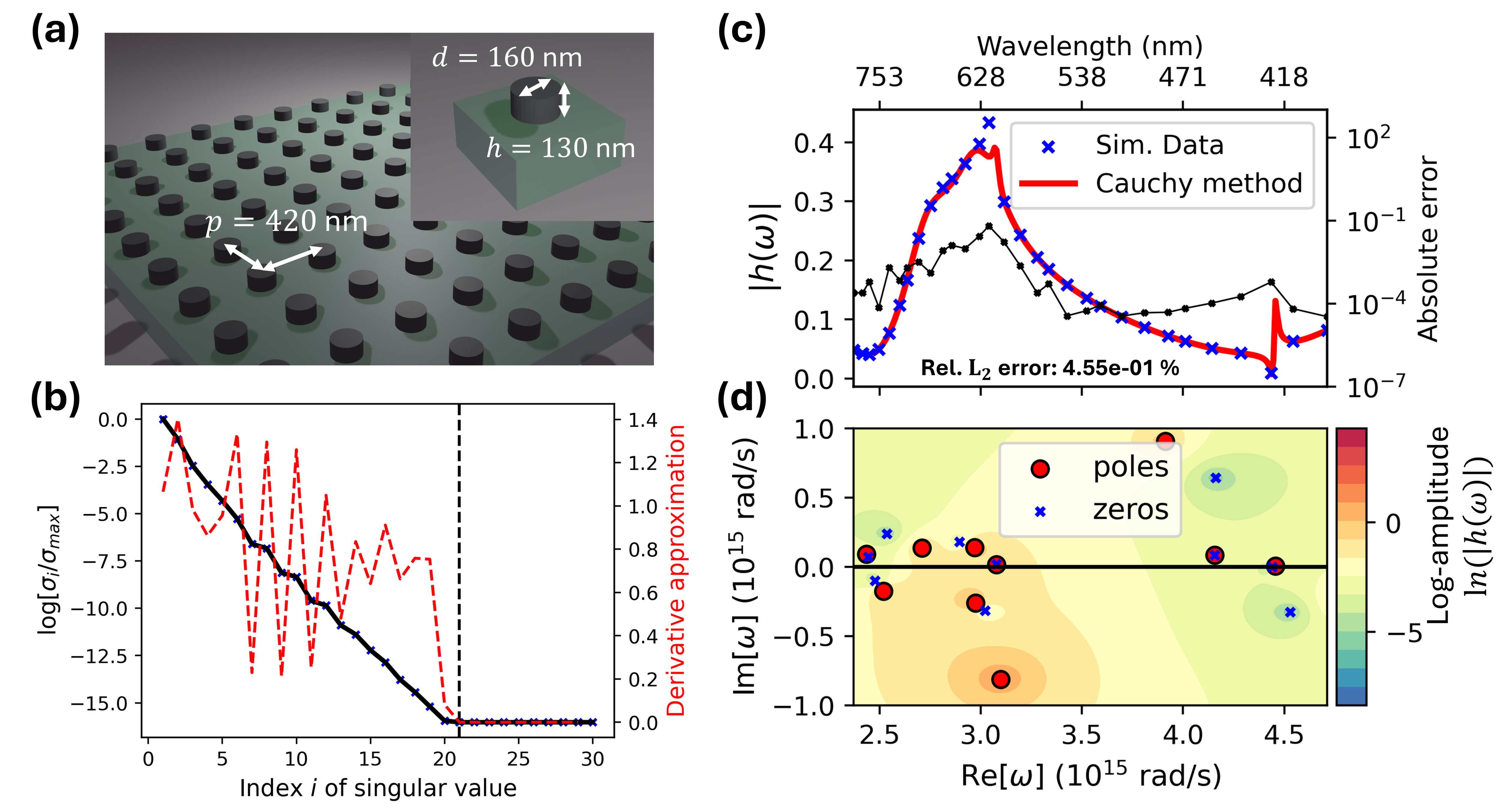}
    \caption{(a) Si nanodisks of diameter $d=160$ nm and height $h=130$ nm, arranged in a 2D array of  period $p=420$ nm over a substrate of glass. The signal $\mathbf{h}$ is the $0^{th}$-order optical reflection coefficient at normal incidence in air. (b) Relative singular values of the matrix $\mathbf{C}=[\mathbf{A}~ -\mathbf{B}]$ with $\mathbf{A}$ and $\mathbf{B}$ defined in Equations~\eqref{eq:5_8} and \eqref{eq:5_9}.  The singular values are sorted by decreasing value, and divided by the first one, $\sigma_{max}$. In the classical Cauchy method, the rank $r$ is set to the index of the smallest singular value after which no significant variation of the ratio $\sigma_i/\sigma_{max}$ is observed. (c) Reflection coefficient $h(\omega)=\tilde{r}_{00}(\omega)$ retrieved \via the classical Cauchy method (red curve), and compared to simulated data (blue markers). The absolute error $|h_i - \hat{h}_i|$, with $h_i$ the target and $\hat{h}_i$ the reconstructed value, is plotted (black curve), and shows the high accuracy of the method, with a relative $L_2$ error $e^{(2)}_{22,11,10}=4.55\times 10^{-1}$\%. (d) Distribution of poles and zeros retrieved by the Cauchy method in the complex $\omega$ plane. Some poles and zeros are located outside the spectral window of interest.}
    \label{fig:5_1}
\end{figure}

\noindent
The Cauchy method first requires an estimation of the rank $r$. This is achieved by first choosing large values of $M_p$ and $M_z$, and then performing a Singular Value Decomposition (SVD) of $\mathbf{C}$. The rank $r$ corresponds to the index of the smallest singular value not considered as null, when all the singular values are sorted in descending order.\\

\noindent
We illustrate this process in an example. The response function $h(\omega)$ is the $0^{th}$ order reflection coefficient $h(\omega)=r_{00}(\omega)$ of a 2D array of Si nanodisks illustrated in Figure~\ref{fig:5_1} (a). The initial numbers of poles and zeros are $M_p=20$ and $M_z=19$. The first null singular value is identified by the inflexion point in Figure~\ref{fig:5_1} (b), which displays the sorted relative singular values.\\

\noindent
Once the rank $r$ is estimated, we can select $M_z$ and $M_p$ in order to have $K=1$ in Equation~\eqref{eq:5_10}. Let us point out that, classically, $M_z$ is equal to $M_p-1$ in the Cauchy method.  After these values of $M_z$ and $M_p$ have been selected, we can calculate one solution to Equation~\eqref{eq:5_7} by first performing a QR decomposition of $\mathbf{A}$:
\begin{equation}
    \mathbf{A} = \mathbf{Q} \begin{bmatrix}
        \mathbf{R}_{11} \\
        0 \\
    \end{bmatrix}
\label{eq:5_11}
\end{equation}
with $\mathbf{Q}$ a $N\times N$ orthogonal matrix and $\mathbf{R}_{11}$ a full-rank $(M_z+1)\times(M_z+1)$ upper triangular matrix. By applying $\mathbf{Q}^T$ to $\mathbf{C}$, we obtain
\begin{equation}
    \begin{bmatrix}
        \mathbf{R}_{11} & \mathbf{R}_{12} \\
        0 & \mathbf{R}_{22} \\
    \end{bmatrix}
    \begin{bmatrix}
        \mathbf{a}\\
        \mathbf{b}\\
    \end{bmatrix} = 0 
\label{eq:5_12}
\end{equation}
In Equation~\eqref{eq:5_12}, $\mathbf{b}$ must satisfy
\begin{equation}
    \mathbf{R}_{22} \mathbf{b} = \mathbf{0}.
\end{equation}
The vector $\mathbf{b}$ is thus set to the eigenvector associated with the $0$ eigenvalue of $\mathbf{R}_{22}$, which is uniquely defined up to a multiplicative constant since $K=1$. We can then obtain $\mathbf{a}$ from Equation~\eqref{eq:5_12}:
\begin{equation}
    \mathbf{a} = -\mathbf{R}_{11}^{-1} \mathbf{R}_{12} \mathbf{b}.
\label{eq:5_13}
\end{equation}
Knowing the polynomial coefficients $\mathbf{a}$ and $\mathbf{b}$, $f(\omega)$ and $g(\omega)$ can be calculated, and the SZF expression can be retrieved. The poles $p^{(\ell)}$ are obtained by calculating the roots of $g(\omega)$, and the zeros $z^{(\ell)}$ through the roots of $f(\omega)$. The Cauchy method thus provides a mean to reconstruct a signal through the SZF.\\

\noindent    
We test the classical Cauchy method and show its accuracy in Figure~\ref{fig:5_1} (c,d), using the reflection coefficient $h(\omega)=r_{00}(\omega)$ introduced in Figure~\ref{fig:5_1} (a). The response function is evaluated at $30$ frequencies ranging from $2.3$ to $4.7 \times 10^{15}$ rad/s. The value of the rank is $r=22$, resulting in $M_p=11$ and $M_z=M_p-1=10$. Despite the small numbers of poles and zeros, we observe a good fitting over a large spectral window, with a relative $L_2$ error $e^{(2)}_{21,11,10}$ of $4.55\times 10^{-1}$\%, defined as
\begin{equation}
    e^{(2)}_{r,M_p,M_z}\left[\hat{\mathbf{h}}, \mathbf{h}\right] = \frac{{||\hat{\mathbf{h}} - \mathbf{h}||}_2}{{||\mathbf{h}||_2}}
\label{eq:5_14}
\end{equation}
where $\mathbf{h}=\left(h_1, ..., h_N\right)$ is the experimental data vector, and $\hat{\mathbf{h}}=\left(\hat{h}_1, ..., \hat{h}_N\right)$ is the reconstructed response vector using the parameters $r$, $M_p$ and $M_z$. The norm of a vector $\mathbf{v}=\left(v_1, ..., v_N\right)$ reads as
\begin{equation}
    {||\mathbf{v}||}_2 = \sqrt{\sum_{i=1}^N |v_i|^2}.
\label{eq:5_15}
\end{equation}

\subsection{Accuracy-driven Cauchy method}

\noindent
While the classical Cauchy method provides good results in most situations, it can be improved by modifying some steps, in particular the calculation of the rank and the determination of the final numbers of poles and zeros. In the classical method, the rank is estimated by performing an SVD on $\mathbf{C}$, and is given by the index of the last non-null singular value, as previously shown in Figure~\ref{fig:5_1} (b). Consequently, there is no guarantee that the dimension $K$ of the kernel will be equal to $1$. In addition, there is usually no \apriori information imposing $M_z=M_p-1$. The accuracy of the Cauchy method can thus be easily enhanced by enforcing a kernel dimension $K$ of $1$, and by sweeping through different values of $M_p-M_z\geq0$.\\

\noindent
We first choose a large initial number $M_{p,0}$ of poles, and set the initial number of zeros $M_{z,0}$ to that same value. This results in a matrix $\mathbf{C}_0 = [\mathbf{A}_0, -\mathbf{B}_0]$ of dimension $N \times 2(M_{p,0}+1)$, with $\mathbf{A}_0$ and $\mathbf{B}_0$ defined in Equations~\eqref{eq:5_8} and ~\eqref{eq:5_9}. We then perform an SVD of $\mathbf{C}_0$, and use the same method as described in Figure~\ref{fig:5_1} (b) to obtain a value $r_{\max}$ instead of an approximation of the rank this time. If $r_{\max}$ is even, we add $1$ to get an odd number.\\

\noindent
This leads to two new maximum numbers of poles and zeros $M_{p,\max}=M_{z,\max}=r_{\max}/2$. They are used as boundaries as we iteratively increase the numbers of zeros $M_z$ and poles $M_p$:
\begin{gather}
    1\leq M_z \leq M_{z,\max}\\
    M_z+\Delta D \leq M_p \leq M_{p,\max}.
\end{gather}
$\Delta D$ is a free parameter of the method corresponding to the maximum difference between the number of poles and the number of zeros. A total of $\Lambda=M_{z,\max} \times (M_{z,\max} - \Delta D)$ couples $(M_z, M_p)$ are tested. These combinations are identified by the index $\lambda$, \ie $1\leq \lambda \leq \Lambda$.\\

\noindent
For each value of $\lambda$, a matrix $\mathbf{C}_\lambda$ is calculated, defined once again through Equations~\eqref{eq:5_8} and ~\eqref{eq:5_9}. We can then perform another SVD:
\begin{equation}
    \mathbf{C}_\lambda = \mathbf{U} \begin{bmatrix}
        \mathrm{\mathbf{diag}(\sigma_1, ..., \sigma_{r+1})}\\
        0 \\
    \end{bmatrix} \mathbf{V}^H
\end{equation}
where $\mathbf{U}$ and $\mathbf{V}$ are two unitary matrices, $\mathbf{V}^H$ is the conjugate transpose of $\mathbf{V}$, $r=M_z+M_p+2$, and $\sigma_1, \sigma_2, ..., \sigma_{r_+1}$ are the singular values of $\mathbf{C}_\lambda$. By setting $\sigma_{r+1}$ to $0$, we enforce $r$ as the rank of the resulting matrix $\hat{\mathbf{C}}_\lambda$, ensuring a dimension of the kernel $K=1$. We then perform the classical Cauchy method on all matrices $\hat{\mathbf{C}}_\lambda$, resulting in $\Lambda$ solutions $[\mathbf{a}_\lambda, \mathbf{b}_\lambda]$, each associated with a reconstruction $\hat{\mathbf{h}}_\lambda$ using Equations~\eqref{eq:5_3} to ~\eqref{eq:5_5}.\\

\begin{figure}[ht!]
    \centering
    \includegraphics[width=.8\textwidth]{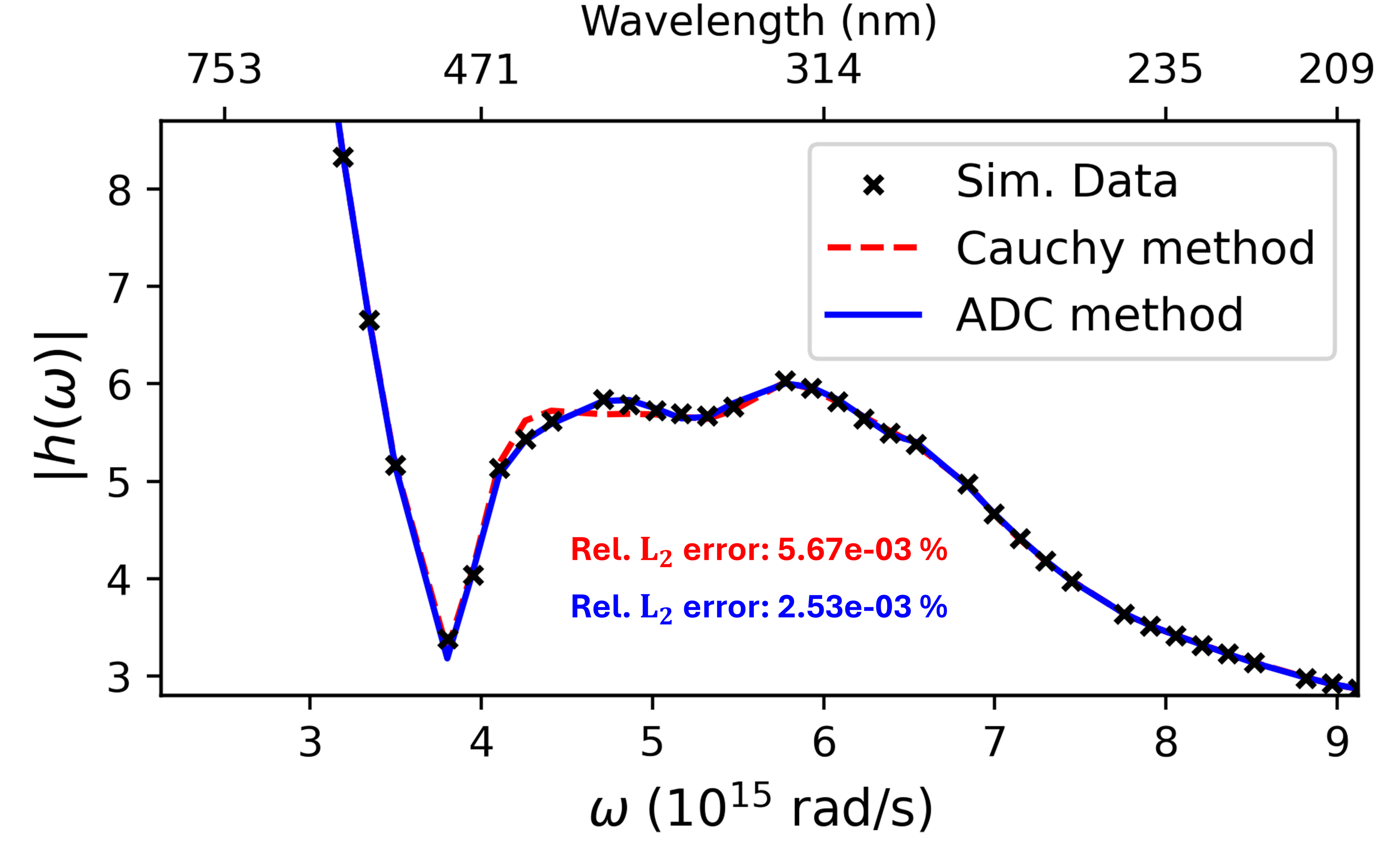}
    \caption{Reconstruction of the relative permittivity $\varepsilon_{\mathrm{Au}}(\omega_i)=h(\omega)$ of $\mathrm{Au}$, using the classical Cauchy method (dashed, red curve) and the ADC method (blue curve). A better reconstruction is obtained with the ADC method, which yields a relative $L_2$ error $e^{(2)}_{18,9,8}=2.53\times 10^{-3}$\%, lower than that obtained \via the classical Cauchy method, which is $e^{(2)}_{20,10,9}=5.67\times 10^{-3}$\%.}
    \label{fig:5_2}
\end{figure}

\noindent
The relative $L_2$ error $e^{(2)}_{r,M_p,M_z}\left[\mathbf{h}_\lambda, \mathbf{h}\right]$ corresponding to each solution is evaluated. The optimal solution $\mathbf{h}_{\lambda^*}$ is chosen, defined as
\begin{equation}
    \lambda^* = \argmin_{\lambda} e^{(2)}_{r,M_p,M_z}\left[\mathbf{h}_\lambda, \mathbf{h}\right].
\end{equation}
We refer to this enhanced approach as the Accuracy-Driven Cauchy method (ADC method). The ADC method is, by construction, always at least as accurate as the classical method when the same initial amount of poles $M_{p,0}$ is chosen. We compare the two approaches in Figure~\ref{fig:5_2}, where the response function is the relative permittivity of gold $h(\omega)=\varepsilon_{\mathrm{Au}}(\omega)$, and assess the increased performance of the ADC method compared to the conventional Cauchy method.

\subsection{Physics-informed corrections}

\noindent
So far, we have focused on the quality of the reconstruction without paying attention to the distributions of poles and zeros. Nevertheless, these complex frequencies are constrained in the case of physical systems and signals. We propose a means to take into account the properties of physical systems while maintaining an accurate fitting of the data.\\ 

\noindent
If $h(\omega)$ is the response or the transfer function of a physical system, \revision{we generally assume that it is the Fourier-transform of a real-valued function in the temporal domain. As a consequence, $h(\omega)$ is Hermitian-symmetric, \ie $h(\omega)=\overline{h(-\overline{\omega})}$. Its} poles and zeros come in pairs $(p^{(\ell)}, -\overline{p^{(\ell)}})$ and $(z^{(\ell)}, -\overline{z^{(\ell)}})$, and the constant $\eta_0$ is real-valued~\cite{ben_soltane_multiple-order_2023}. We can rewrite the truncated SZF as
\begin{equation}
    \frac{h(\omega)}{\eta_0} = \frac{\displaystyle \prod_{\ell=1}^{M_{z,\rmi\mathbb{R}}}(\omega - \rmi z_{\rmi \mathbb{R}}^{(\ell)})}{\displaystyle \prod_{\ell=1}^{M_{p,\rmi\mathbb{R}}}(\omega + \rmi p_{\rmi \mathbb{R}}^{(\ell)})} \frac{\displaystyle \prod_{\ell=1}^{M_{z,\mathbb{C}}}\left[(\omega - z_{\mathbb{C}}^{(\ell)})(\omega + \overline{z_{\mathbb{C}}^{(\ell)}})\right]}{\displaystyle \prod_{\ell=1}^{M_{p,\mathbb{C}}}\left[(\omega - p_{\mathbb{C}}^{(\ell)})(\omega + \overline{p_{\mathbb{C}}^{(\ell)}})\right]}
\label{eq:5_16}
\end{equation}
where $\rmi z_{\rmi \mathbb{R}}$ are purely imaginary zeros, $-\rmi p_{\rmi \mathbb{R}}$ are purely imaginary poles, $z_{\mathbb{C}}$ are non-imaginary zeros, and $p_{\mathbb{C}}$ are non-imaginary poles. The number of poles and zeros read as $M_z=M_{z,\rmi \mathbb{R}} + M_{z, \mathbb{C}}$ and $M_p=M_{p,\rmi \mathbb{R}} + M_{p, \mathbb{C}}$.\\

\noindent
In order to take into account this Hermitian symmetry, we add negative frequencies, effectively multiplying the number of frequencies by 2. For every frequency $\omega_i$ associated with the value $h_i$, the opposite frequency $-\omega_i$ is added, along with $\overline{h_i}$. This operation is performed at the beginning of the Cauchy method, right before the step corresponding to Equation~\eqref{eq:5_6}.\\

\noindent
The resulting distributions of poles and zeros are not perfectly Hermitian-symmetric, whether we use the ADC method or the classical Cauchy method. We correct for discrepancies by removing all the poles and zeros with a negative real part, before replacing them with the Hermitian-symmetrics of the poles and zeros with a positive real part. We thus explicitly impose the symmetry expressed in Equation~\eqref{eq:5_16} to reconstruct $h(\omega)$. In the ADC method, the poles and zeros are replaced before calculating the errors $e^{(2)}_{r,M_p,M_z}[\mathbf{h}_\lambda, \mathbf{h}]$ in order to ensure that the optimal Hermitian-symmetric solution is chosen among all the calculated options.\\

\noindent
In addition, the poles and zeros located too far from the window of interest act as offsets and have little influence over the variations of $h(\omega)$ at real frequencies. We thus remove them from the sets of poles and zeros, but modify the constant term $\eta_0$ in order to take into account their contributions:
\begin{equation}
    \eta_0' = \eta_0 \frac{\displaystyle \prod_{z_f}\left(-|z_f|^2\right)}{\displaystyle \prod_{p_f}\left(-|p_f|^2\right)} 
\label{eq:5_17}
\end{equation}
where $\eta_0'$ is the modified constant term, and $z_f$ and $p_f$ are the zeros and poles far from the spectral range of interest. They are considered as such when their distance to the origin exceeds a certain threshold $\rho$, which is by default set to $5$ times the spectral range. The calculation of $\eta_0'$ is also performed before the calculation of the errors in the ADC method.\\

\noindent
Finally, we propose to further modify the distribution of poles to account for the usually assumed stability of physical systems or signals. \revision{When an inverse Fourier-transform is performed to return to the temporal domain, each pole $p^{(\ell)}$ with residue $r^{(\ell)}$ gives rise to a complex exponential term $-\rmi e^{-\rmi\Real[p^{(\ell)}]t}e^{+\Imag[p^{(\ell)}]t}$, which grows to $+\infty$ in modulus if $\Imag[p^{(\ell)}]>0$~\cite{colom_modal_2018,ben_soltane_multiple-order_2023}. A stable or non-diverging system or signal must thus possess poles with a negative imaginary part. By extension, a pole is called stable if its imaginary part is negative. To favour stable solutions,} we penalize the solutions associated with unstable poles in the ADC method by multiplying the errors $e^{(2)}_{r,M_p,M_z}\left[\mathbf{h}_\lambda, \mathbf{h}\right]$ by $(1+M_{inst,p})$ with $M_{inst,p}$ the number of poles with a strictly positive imaginary part. What is more, we add a negative imaginary part $-\rmi q_0$ to the poles located on the real frequency axis once the optimal solution is obtained. $q_0$ is yet another free parameter set to $10^{-5}$ by default. Since the effect of this modification on the zeros is not easily obtained, this last step is performed after converting the SZF into the SEM, \ie by considering the poles and their associated residues instead of considering the poles and zeros.\\

\noindent
The residue $r^{(m)}$ associated with a pole $p^{(m)}$ is calculated as
\begin{equation}
    r^{(m)} = \eta_0 ~ \frac{\displaystyle \prod_{\ell=1}^{M_z}(p^{(m)} - z^{(\ell)})}{\displaystyle \prod_{\ell \neq m}(p^{(m)} - p^{(\ell)})}
\end{equation}
and the non-resonant term is obtained through the expression~\cite{ben_soltane_derivation_2022}
\begin{equation}
    h_{NR} = h(\omega_0) + \sum_{\ell=1}^{M_p} \frac{r^{(\ell)}}{p^{(\ell)}}
\end{equation}
where $\omega_0$ is an arbitrary complex frequency different from the poles. In addition, we remove the poles associated with residues which have a small modulus compared to the maximum modulus. The default threshold is $1 \%$ of the maximum.

\begin{figure}[ht!]
    \centering
    \includegraphics[width=.78\textwidth]{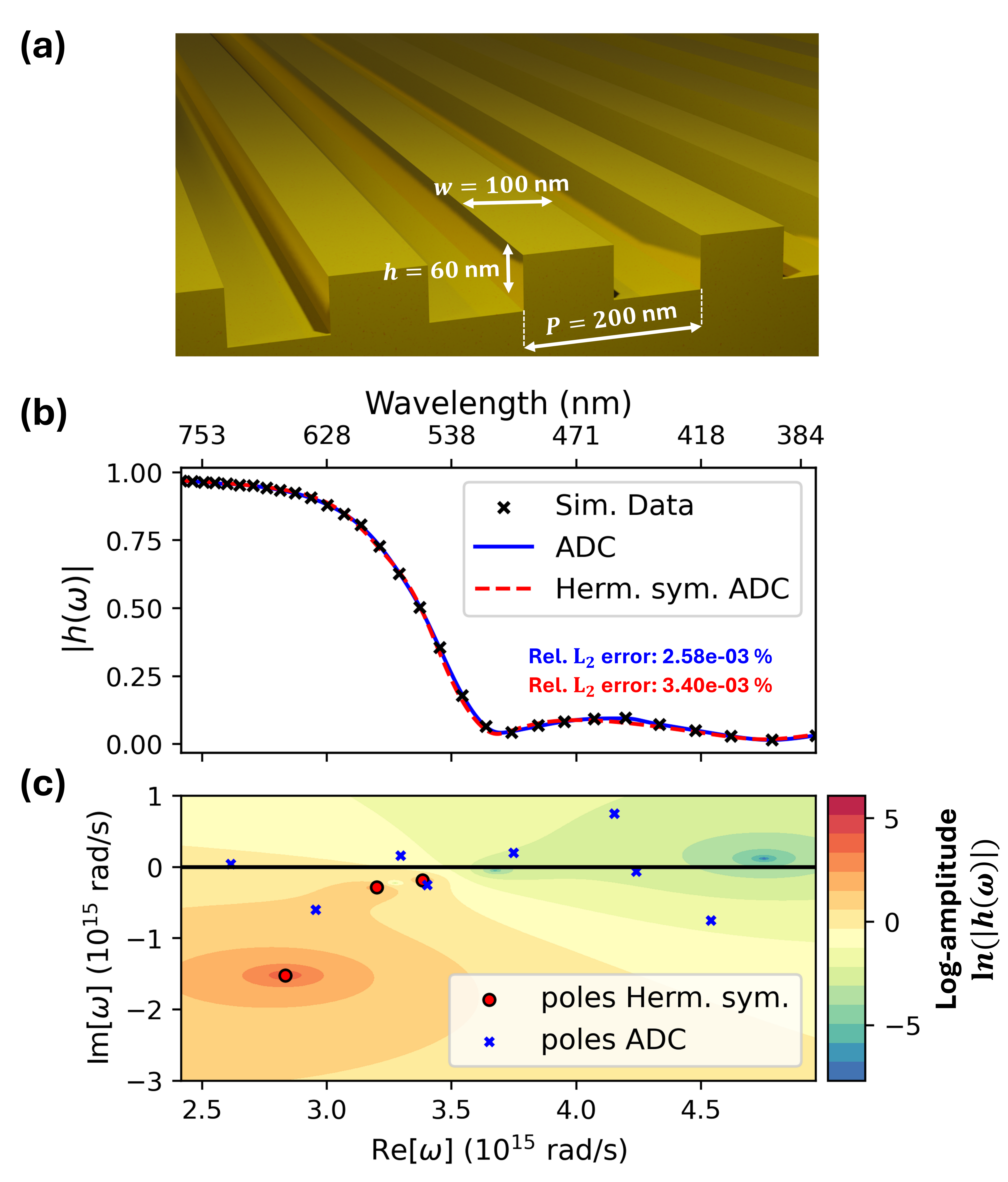}
    \caption{(a) 1D gold grating over a substrate of gold, with height $h=60$ nm, width $w=100$ nm, and period $p=200$ nm. The studied function is the $0^{th}$ order reflection coefficient $r_0(\omega)=h(\omega)$ at normal incidence in air. (b) Approximation of $h(\omega)$, using the ADC method (blue curve, ADC) or the physics-informed version (red curve, Herm. sym. ADC). (c) Distributions of poles in the complex $\omega$ plane using the two approaches. While both versions provide similar results, \ie an accurate fitting of the experimental data (black markers), the physics-informed method fulfils the Hermitian symmetry and requires fewer pairs of poles and zeros.}
    \label{fig:5_3}
\end{figure}

\noindent
We compare the ADC method to its Hermitian-symmetric counterpart in the case of the reflection coefficient of a 1D gold grating described in Figure~\ref{fig:5_3} (a). We show that the reconstruction is only slightly affected by the proposed changes in Figure~\ref{fig:5_3} (b, c).\\

\noindent
The Cauchy method may fail to provide accurate models in the case of rapidly varying curves, or when stability is mandated, as it will be shown in the last section of this manuscript. Despite its rapidity of execution, it must therefore be replaced in these cases by an alternative method, which is offered here by the auto-differentiation approach.

\section{Auto-differentiation approach for the singularity expansion}

\subsection{Formulating the optimization problems}

\noindent
We now move onto the second approach which aims at exploiting auto-differentiation (as proposed in the PyTorch library) to obtain accurate approximations through optimization problems. Although these approaches could be applied to optimize the SZF, we focus on the retrieval of the different terms in the SEM, or in other words, the poles $p^{(\ell)}$, the residues $r^{(\ell)}$, and the non-resonant term $h_{NR}$. Similarly to the previous section, we consider the truncated SEM expression, with $M$ complex pairs of poles or purely imaginary poles:
\begin{equation}
    h(\omega) = h_{NR} + \sum_{\ell=1}^{M} \left[ \frac{r^{(\ell)}}{\omega - p^{(\ell)}} - \frac{\overline{r^{(\ell)}}}{\omega + \overline{p^{(\ell)}}} \right].
\label{eq:5_24}
\end{equation}
While it is possible to deal with complex values in the optimization process, current auto-differentiation tools have only recently been able to deal with such numbers, and some functions are still not properly defined in $\mathbb{C}$~\cite{chu2021scheme,kramer2024tutorial}. We thus express the parameters in $\mathbb{R}^2$ instead of $\mathbb{C}$.\\

\revision{
\noindent
The parameters, \ie the real and imaginary parts of the poles $p^{(\ell)}$ and residues $r^{(\ell)}$, and the real-valued non-resonant term $h_{NR}$, are obtained by minimizing a loss function \via a gradient-descent(-like) method.} The loss function that we use requires the definition of the relative $L_2$ and $L_\infty$ errors:
\begin{equation}
    e_{q}\left[\mathbf{h}, \hat{\mathbf{h}}\right] = \frac{{||\mathbf{h} - \hat{\mathbf{h}}||}_q}{{||\mathbf{h}||_q}}
\label{eq:5_28}
\end{equation}
with $q\in\{2, \infty\}$, \ie
\begin{equation}
    \begin{gathered}
        {||\mathbf{v}||}_2 = \sqrt{\sum_{i=1}^N |v_i|^2} \\ 
        {||\mathbf{v}||}_\infty = \max_i |v_i|.
    \end{gathered}
\label{eq:5_30}
\end{equation}
The loss function itself is then defined as
\begin{equation}
    \mathcal{L}[\mathbf{h}, \hat{\mathbf{h}}] =
    \alpha_1 e^{(2)}\left[\mathbf{h}, \hat{\mathbf{h}}\right] + 
    \alpha_2 \norm{ \frac{\mathbf{h} - \hat{\mathbf{h}}}{\mathbf{h}}}_\infty + \alpha_3\left<\frac{\delta \mathbf{h}_R}{\mathbf{h}_R}\right> + \alpha_4\left<\frac{\delta \mathbf{h}_I}{\mathbf{h}_I}\right>
\label{eq:5_31}
\end{equation}
where $\mathbf{h}$ is the target, $\hat{\mathbf{h}}$ is the prediction, and the notation $<.>$ corresponds to the averaging operation over all the sampled frequencies. $\delta \mathbf{h}_R$ is the vector of errors between the real part of the model $\hat{\mathbf{h}}$ and the experimental data $\mathbf{h}$, and $\mathbf{h}_R$ is the vector of the absolute values of the real part $\mathbf{h}$. It is used to change the weight given to the frequency $\omega_i$ in $\mathbf{\delta h_R}$. $\mathbf{\delta h_I}$ and $\mathbf{h}_I$ do the same for the imaginary part:
\begin{align}
    \delta h_{R,i} &= \left|\mathrm{Re}[h_i - \hat{h}_i]\right|, & h_{R,i} &= \left|\mathrm{Re}[h_i]\right| + .5, \label{eq:5_32}\\
    \delta h_{I,i} &= \left|\mathrm{Im}[h_i - \hat{h}_i]\right|, & h_{I,i} &= \left|\mathrm{Im}[h_i]\right| + .5. \label{eq:5_33}
\end{align}

\noindent
The values of the weight coefficients are usually case-dependent. The first coefficient $\alpha_1$ roughly corresponds to the average error between $\mathbf{h}$ and $\hat{\mathbf{h}}$. $\alpha_2$ can be used to give more weight to small variations in the amplitude. $\alpha_3$ and $\alpha_4$ also allow to strengthen the contribution of points associated with small variations or values, but in the real and imaginary parts respectively.\\

\begin{figure}[ht!]
    \centering
    \includegraphics[width=.9\textwidth]{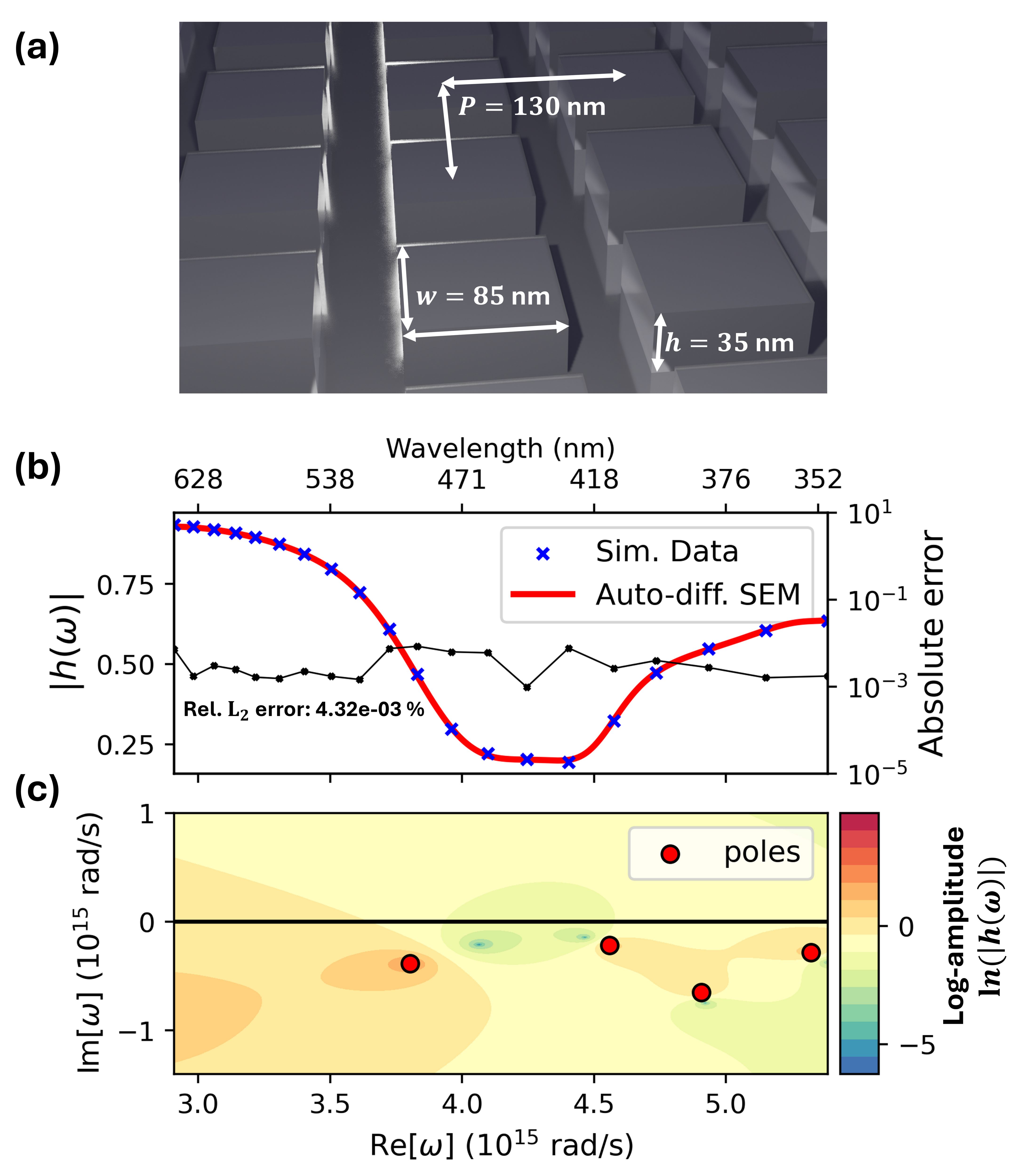}
    \caption{(a) 2D array of square Ag pillars of width $w=85$ nm and height $h=35$ nm, and period $p=130$ nm over a substrate of Ag. (b) The signal is the $0$-th order reflection coefficient $h(\omega)=r_{00}(\omega)$ of the 2D array depicted in (a) illuminated in normal incidence reconstructed with the SEM in Equation~\eqref{eq:5_24} (red curve), and compared to simulated data (blue markers). The absolute error $|h_i - \hat{h}_i|$ (black curve) is less than $10^{-3}$ at all frequencies, and a relative $L_2$ error of $4.32\times 10^{-3}$ \% is obtained. (c) Distribution of poles in the complex $\omega$ plane.}
    \label{fig:5_4}
\end{figure}

\noindent
\revision{Using the example of the $0$-th order reflection coefficient of a 2D grating of silver described in Figure~\ref{fig:5_4} (a), we show the result of the retrieval of the singularity expansion of Equation~\eqref{eq:5_24} in Figure~\ref{fig:5_4} (b,c), with $M_I=0$ imaginary poles, $M_C=7$ complex pairs of poles, and $\mathbf{\alpha}=(1,0,0.2,0.2)$. The non-imaginary poles were initialized with their real parts evenly distributed from $2.5$ to $4.9\times 10^{15}$ rad/s, and their imaginary part set to $-0.05$ times their real parts. We obtain a relative $L_2$ error of $4.32\times 10^{-3}$\%, with a Hermitian-symmetric solution.}

\subsection{Other forms of the singularity expansion}

\noindent
We can cast the SEM expression into different forms which highlight different properties of physical systems: 
\begin{itemize}
    \item The classical SEM expression in Equation~\eqref{eq:5_1_sem} exhibits resonant terms associated with singularities and residues in the complex $\omega$ plane, represented as peaks, and intuitively introduces the analytical continuation of $h(\omega)$ in the complex plane.
    \item The Generalized Drude-Lorentz model (GDL)~\cite{ben_soltane_generalized_2024} models the physical systems as sets of sub-units, each behaving as oscillators described by classical mechanics. The sub-units can be identified as the microscopic-scale particles constituting the system.\\
\end{itemize}

\noindent
As shown in reference~\cite{ben_soltane_generalized_2024}, we can recast the singularity expansion into the GDL \via a change of variables for the $M_C$ symmetric terms:
\begin{gather}
    \omega_{0,\ell}=|p_R^{(\ell)}| \label{eq:5_35}\\
    \Gamma_\ell=-2|p_I^{(\ell)}| \label{eq:5_36}\\
    s_{1,\ell} \Gamma_l = -2b^{(\ell)} \label{eq:5_37}\\ 
    s_{2,\ell} \left(\omega_{0,\ell}^2 + \frac{\Gamma_\ell}{2}^2 \right) = -2\Real\left(r_p^{(\ell)}~\left(\omega_{0,\ell} - \rmi\frac{\Gamma_\ell}{2} \right)\right) \label{eq:5_38}
\end{gather}
and by rewriting the $M_I$ resonant terms associated with imaginary poles as
\begin{gather}
    \gamma_\ell = -q^{(\ell)} \label{eq:5_39}\\   
    \omega_{b,\ell}^2 = |\rho^{(\ell)} \gamma_\ell| \label{eq:5_40}\\
    \gamma_0 = r_0 + \sum_{\ell=1}^{M}\frac{\omega_{b,\ell}^2}{\gamma_\ell} \label{eq:5_41}\\
    \rmi\frac{r_0}{\omega} + \sum_{\ell=1}^{M} \frac{\rho^{(\ell)}}{\omega-\rmi q^{(\ell)}} = \rmi\frac{\gamma_0}{\omega} - \sum_{\ell=1}^{M} \frac{\omega_{b,\ell}^2}{\omega^2 + \rmi\omega\gamma_\ell} \label{eq:5_42}
\end{gather}
where we assume that there is a pole $0$ at the origin, associated with a residue $\rmi r_0$. 

\begin{figure}[H]
    \centering
    \includegraphics[width=.9\textwidth]{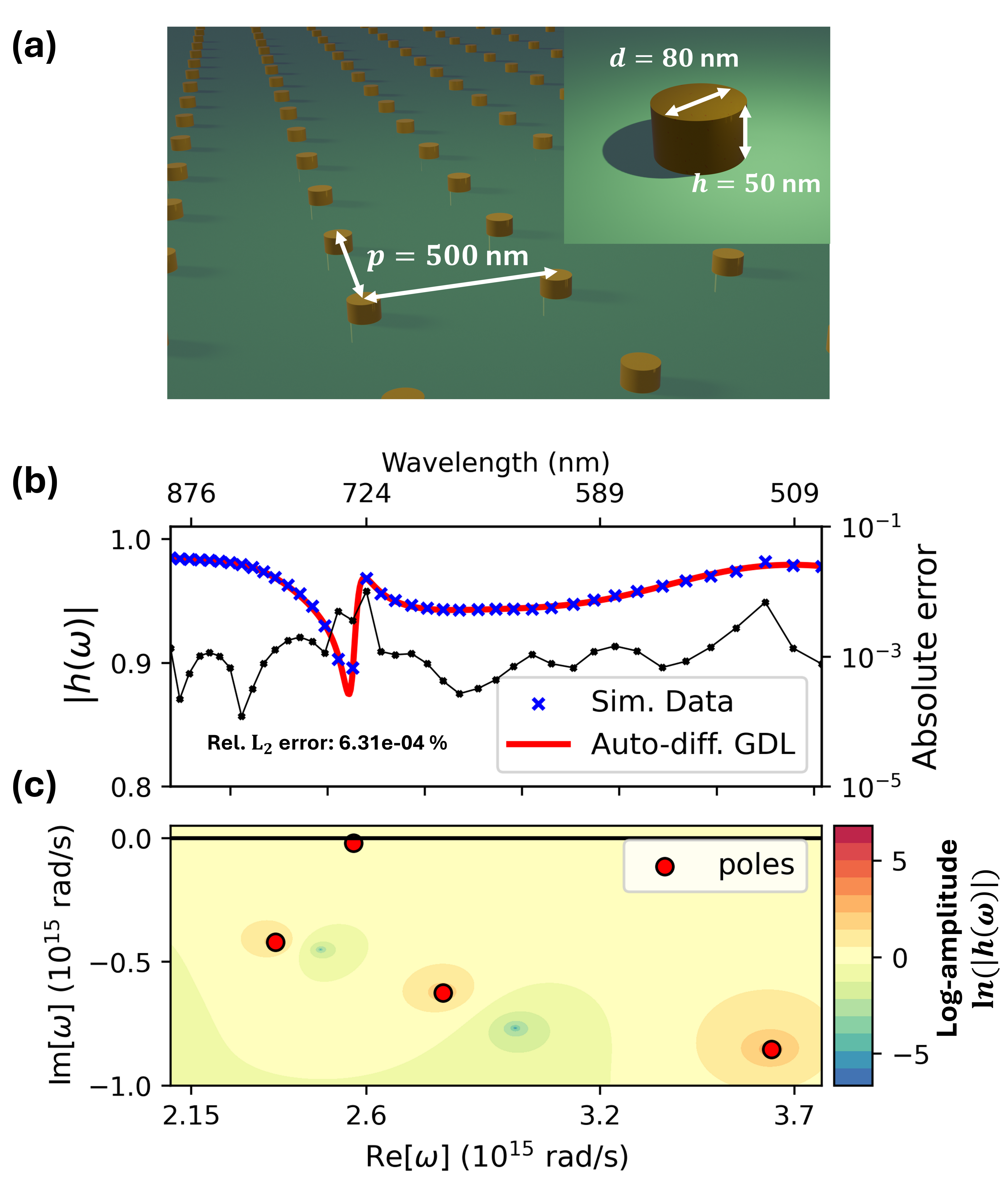}
    \caption{(a) Lattice of gold nanodisks with height $h=50$nm, diameter $d=80$nm and period $p=500$ nm, bearing on a substrate of glass. The signal is the $0^{th}$-order transmission coefficient $h(\omega)=t_{00}(\omega)$ for a normal incidence in air. (b) Reconstruction of $h(\omega)$ with the GDL in Equation~\eqref{eq:5_43} (red curve), and compared to simulated data (blue markers). A relative $L_2$ error $e^{(2)}=6.31 \times 10^{-3}$ \% is obtained. (c) Distribution of the poles in the complex $\omega$ plane.}
    \label{fig:5_5}
\end{figure}

\noindent
Let us stress that the absolute value "$|.|$" is used to impose the stability of all the poles, but it can be removed to allow for unstable effective poles. The following expression of the GDL is obtained:
\begin{equation}
    h(\omega) =  h_{\mathrm{NR}} + \rmi\frac{\gamma_0}{\omega} - \sum_{\ell=1}^{M_I} \frac{\omega_{b,\ell}^2}{\omega^2 + \rmi\omega\gamma_\ell} - \sum_{\ell=1}^{M_C} \frac{\rmi s_{1,\ell}\omega\Gamma_\ell + s_{2,\ell}\left(\omega_{0,\ell}^2 + \frac{\Gamma_\ell}{2}^2 \right)}{\left(\omega^2 - \left(\omega_{0,\ell}^2 + \frac{\Gamma_\ell}{2}^2 \right)\right) + \rmi\omega\Gamma_\ell}.
\label{eq:5_43}
\end{equation}

\noindent
As previously stated, this formulation of the SEM is particularly adapted to the representation of sub-units of a physical system as harmonic oscillators.\\

\noindent
We apply it to the case of the lattice of gold nanodisks presented in Figure~\ref{fig:5_5} (a). The studied signal is the transmission coefficient $h(\omega)=\tilde{t}(\omega)$ at normal incidence. The reconstruction shown in Figure~\ref{fig:5_5} (b,c) is obtained with $M_I=1$ imaginary pole, $M_C=8$ complex pairs of poles and $\mathbf{\alpha}=(1,0,0,0)$. The weight coefficients are such that more importance is given to the real part $h(\omega)$ which shows more complex variations than the imaginary part. We obtain a relative $L_2$ error of $6.31\times 10^{-4}$ \%, over a spectral range extending from $100$nm to $900$nm.

\section{Combined approach}

\noindent
The ADC method and the auto-differentiation approach aim at performing the same task, namely retrieving an analytical expression of a physical signal in the form of the SEM or SZF. Nevertheless, the two methods work in very different ways, both featuring their own limitations and use cases. This makes any direct comparison of the two methods irrelevant.\\

\begin{figure}[ht!]
    \centering
    \includegraphics[width=.8\textwidth]{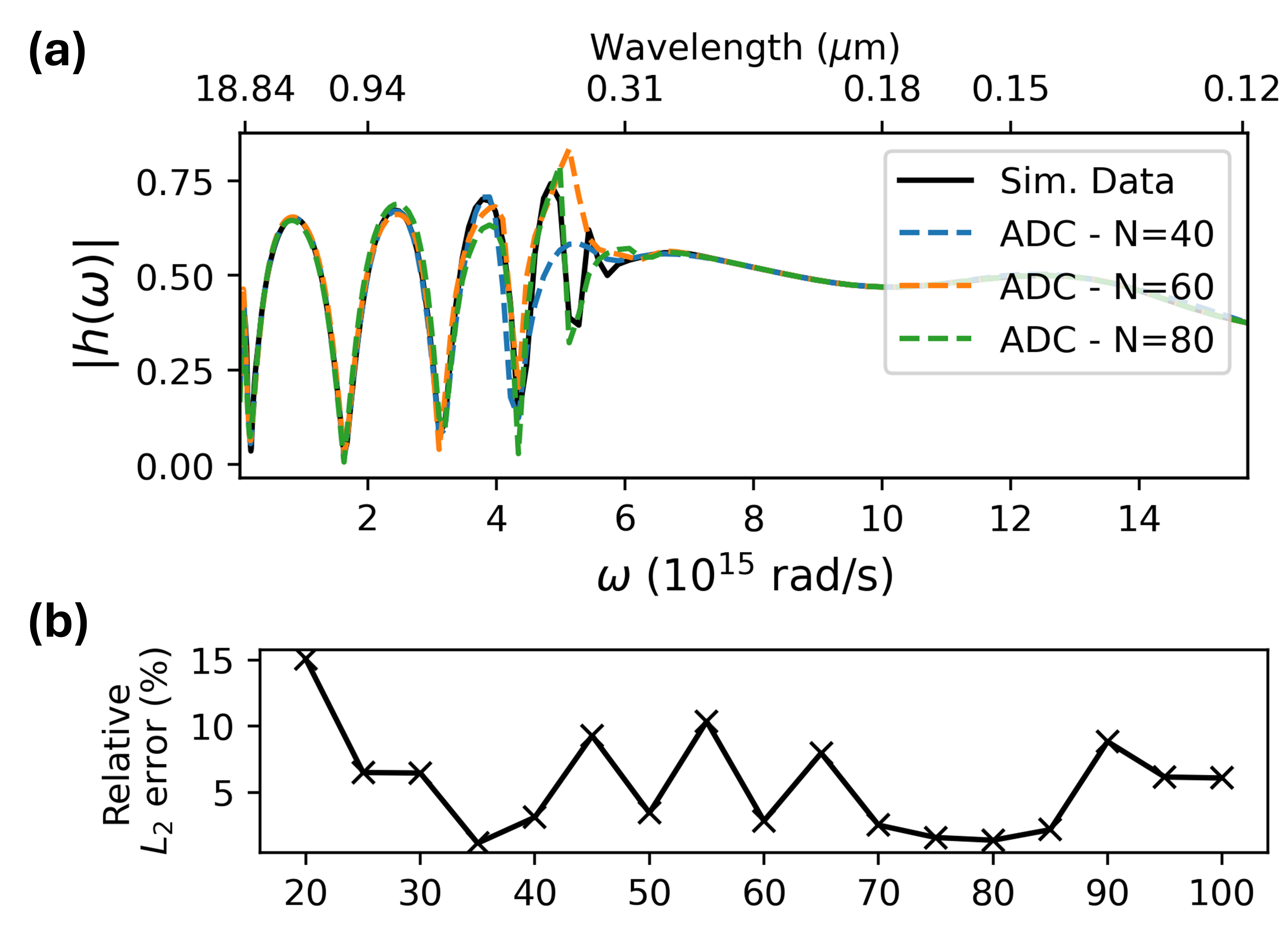}
    \caption{Influence of the frequency sampling in the ADC method on the relative $L_2$ error. The signal is the reflection coefficient of a slab of \tio $h(\omega)=\tilde{r}(\omega)$ of thickness $d=260$ nm, for a TM illumination at $17\degree$ coming from the air superstrate. $N$ frequencies are uniformly picked between $0.15$ and $15.6\times10^{15}$ rad/s. (a) Reconstruction of $h(\omega)$ using $N=40$ (blue), $N=60$ (orange) and $N=80$ (green) frequencies, and compared to the target (black). (b) Evolution of the relative error with the number of frequencies, with $N$ going $5$ by $5$ from $20$ to $100$.}
    \label{fig:5_6}
\end{figure}

\noindent
The ADC method is faster than the auto-differentiation approach. However, it can be difficult to maintain a high accuracy while opting for a Hermitian-symmetric and stable solution, as previously shown. In addition, the method does not work well in large spectral windows showing rapid variations of the signals, and is sensitive to the sampling of data. While the latter is addressed with what is known as the adaptive Cauchy method~\cite{sarkar_application_2000}, the former is not easily solved for. We illustrate both problems in Figure~\ref{fig:5_6}, where we apply the ADC method to retrieve the poles and zeros of the reflection coefficient of a slab of \tio $h(\omega)=\tilde{r}(\omega)$, of thickness $d=260$nm, with a TM illumination at $17\degree$ in air. $N$ frequencies are uniformly picked between $0.15$ and $15.6\times10^{15}$ rad/s. We observe in Figure~\ref{fig:5_6} (a) that fast dynamics such as the one around $5 \times 10^{15}$ rad/s are difficult to capture, whatever the value of $N$. In Figure~\ref{fig:5_6} (b) shows that the relative $L_2$ does not systematically lower when $N$ is increased.\\

\noindent
The auto-differentiation approach can provide accurate solutions even in situations where the ADC method fails. In addition, it allows for an easy implementation of constraints, in particular the Hermitian-symmetry and the stability of poles. However, this efficiency is obtained at the price of a much slower computation time depending on the hyperparameters: the optimization scheme, the desired number of iterations, the learning-rate and the learning-rate updating method, the loss function weight vector $\mathbf{\alpha}$, the sampling of the frequencies $\omega_i$ and values $h_i$, the number of poles and zeros or residues, and their initial distributions.

\begin{figure}[ht!]
    \centering
    \includegraphics[width=.75\textwidth]{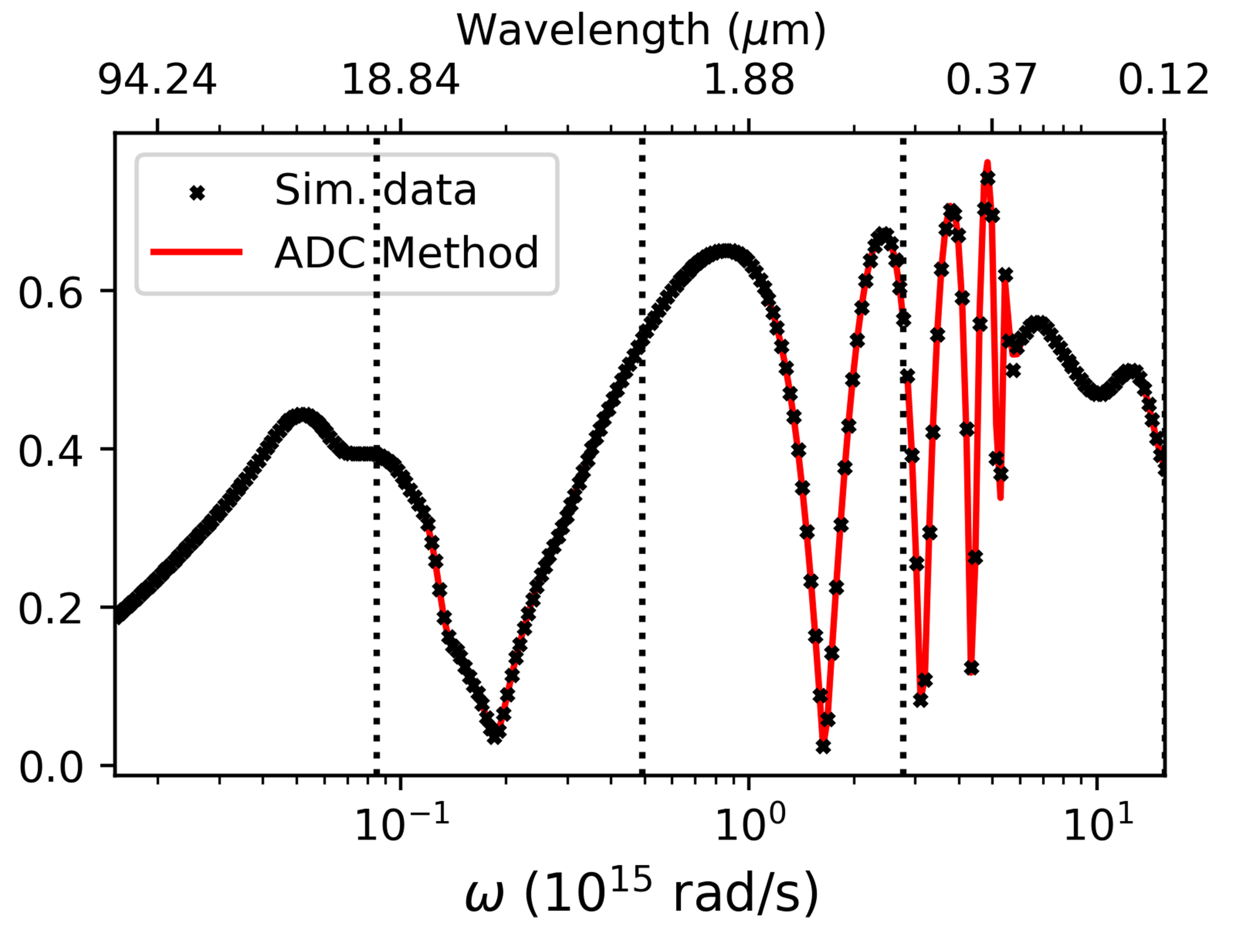}
    \caption{Chained Cauchy method applied to the reconstruction of the reflection coefficient of a slab of \tio of thickness $d=260$ nm, with a TM illumination at $17\degree$ from the air superstrate. The reconstruction is carried out in four smaller spectral windows, which are indicated by the dotted black lines.}
    \label{fig:5_7}
\end{figure}

\noindent
We are particularly interested in the latter. By properly selecting the initial values of the parameters in such a way that the initial approximation is close to the target function, we can reduce the converge time towards a suitable solution, lowering the relative $L_2$ error, and we can often reduce the total number of poles.\\

\noindent
We propose to obtain this initial distribution of poles and zeros through a chained version of the ADC method, which successively fits subsets of the data in smaller frequency windows. In Figure~\ref{fig:5_7}, we show how this applies to $h(\omega)$. We use four sub-windows, each bounded by the vertical dotted lines in Figure~\ref{fig:5_7}. In this case, each window is chosen so that it has the same number of frequency points, \ie $75$ points. Each sub-window is associated with its own sets of poles and zeros and constant $\eta_0$.\\

\noindent
Let us stress that despite the accurate reconstruction obtained for each sub-window, this chained ADC method does not provide a suitable solution on account of three issues:
\begin{enumerate}[label=(\roman*)]
    \item The high number of poles and zeros associated with each sub-window prevents us from discriminating between a fitting of the curve itself and a fitting of the noise. Having too many poles and zeros generally leads to a number of degrees of freedom so high that any curve can be approximated.
    \item While the Hermitian-symmetry can be imposed, it is not the case for the stability criterion. Therefore, the solution is physically irrelevant.
    \item It is not possible to obtain a solution for the whole spectral range directly from the solutions of the sub-windows. We have thus turned one fitting problem into several ones, thereby increasing the amount of required parameters to describe the full response function.
\end{enumerate}
These different problems motivate the use of another method, \ie the auto-differentiation approach, on top of the chained ADC method to obtain a stable physical solution with fewer parameters, among which the relevant ones can be more easily identified.

\begin{figure}[H]
    \centering
    \includegraphics[width=.9\textwidth]{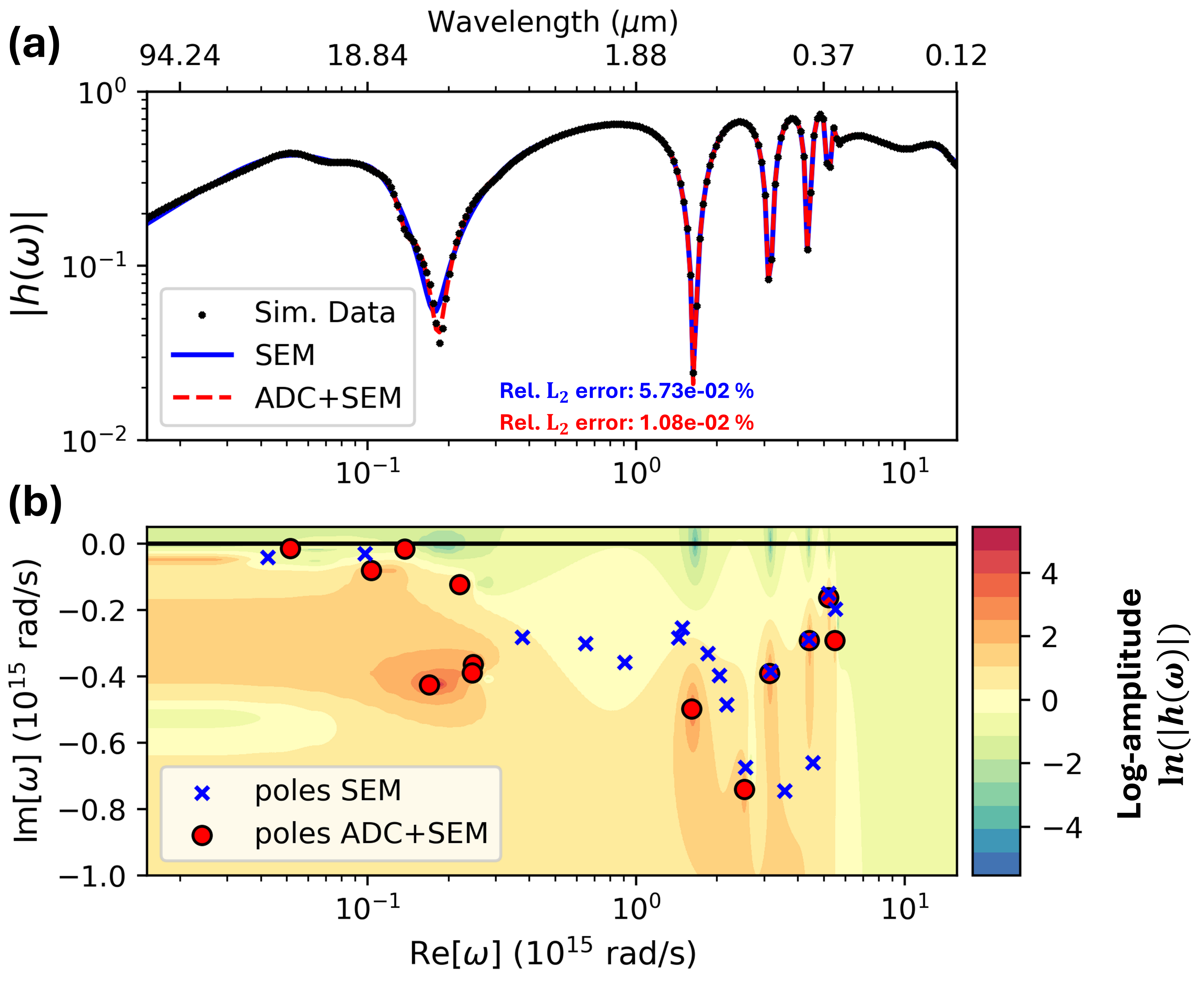}
    \caption{(a) Reflection coefficient $h(\omega)=\tilde{r}(\omega)$ of a slab of \tio of thickness $d=260$ nm, illuminated from the air superstrate with an incidence of $17\degree$, in the TM polarization. It is reconstructed using either the SEM (blue curve), or the combined ADC and SEM approach (red curve), and compared to simulated data (black markers). A relative $L_2$ error $e^{(2)}=5.73 \times 10^{-2}$ \% is obtained with the SEM approach, and $e^{(2)}=1.08 \times 10^{-2}$ \% for the combined approach. The latter thus performs better, and is obtained in half of the time needed for the former. (b) Distribution of the poles in the complex $\omega$ plane, for the SEM (blue markers) and the combined approach (red markers). Fewer singularities are required with the ADC+SEM approach to achieve a better reconstruction than with the SEM.}
    \label{fig:5_8}
\end{figure}

\noindent
For each sub-window, we convert the set of poles and zeros into a set of poles and residues, as well as a constant $h_{NR}$. A contribution weight $q_{w,\ell}$ is defined for all the pole $p^{(\ell)}$, and can be used to filter out poles that do not significantly help in the reconstruction in a sub-window. To do so, we first introduce a score $\rho_{w,\ell}$ which quantifies the variations brought about by the resonant term $h_\ell(\omega)$ associated with $p^{(\ell)}$:
\begin{gather}
    h_\ell(\omega) = \frac{r^{(\ell)}}{\omega - p^{(\ell)}} - \frac{\overline{r^{(\ell)}}}{\omega + \overline{p^{(\ell)}}}\\
    d_m = \min_i |h_p(\omega_i)| \\ 
    d_M = \max_i |h_p(\omega_i)|\\
    \rho_{w,\ell} = 1 - \frac{d_m}{d_M}.
\end{gather}
We also introduce a score $\eta_{w,\ell}$ which quantifies the contribution of the resonant term $h_\ell(\omega)$ to the response function $h(\omega)$ in the sub-window:
\begin{equation}
    \eta_{w,\ell} = \frac{\sum_i |h(\omega_i) - h_\ell(\omega_i)|}{\sum_i |h(\omega_i)|}.
\end{equation}
The weight $q_{w,\ell}$ is subsequently defined as
\begin{equation}
    q_{w,\ell} = \sqrt{\rho_{w,\ell}^2 + \eta_{w,\ell}^2}.
\end{equation}
In every sub-window, the poles with a weight $q_{w, \ell}$ below an arbitrary threshold $q_0=68$ \% are filtered out. We then concatenate all the remaining poles and residues into unique distributions used as initial parameters in the auto-differentiation approach.\\

\noindent
We compare this combined approach to the classical SEM in Figure~\ref{fig:5_8}, for the same reflection coefficient as in Figures~\ref{fig:5_6} and~\ref{fig:5_7}, \ie the case of a slab of \tio. In Figure~\ref{fig:5_8} (a), we observe a slightly more accurate reconstruction with the combined ADC+SEM approach than with the SEM only. Let us stress that the optimization process takes, in this case, half as long with the combined approach than with just the SEM. Another significant advantage is that fewer singularities are needed with the ADC+SEM approach than with the SEM only, as shown in Figure~\ref{fig:5_8} (b), where the singularities associated to both approaches are shown in a frequency window of the complex $\omega$ plane.

\newpage

\section{Pole retrieval efficiency}

\subsection{Performance evaluation criteria}

\noindent
The singularity expansion in Equation~\eqref{eq:5_1_sem} provides an exact expression of any physical signal $h(\omega)$ in terms of the potentially infinite sets of complex poles $p^{(\ell)}$ and associated residues $r^{(\ell)}$. If we assume that $h(\omega)$ is fully known on an open set $\Omega=\left]\omega_0, \omega_1\right[$ of $\mathbb{R}$, then the identity theorem states that there is a unique singularity expansion that matches $h(\omega)$ on $\Omega$, effectively providing an expansion of $h(\omega)$ into the complex frequency plane and ensuring the unicity of the poles, residues and non-resonant term. If we are able to reduce the error between a singularity-expanded function and $h(\Omega)$ to zero, we then obtain the singularity expansion of $h(\omega)$, which we directly denote as $h(\omega)$.

\noindent
Three problems arise in practice.
\begin{enumerate}[label=(\roman*)]
    \item Any prior information is obtained by measuring or simulating $h(\omega)$ over a discrete set of frequencies, \ie the aforementioned $N$ frequencies $\omega_n$. The identity thus no longer holds. Nevertheless, when the spacing between subsequent frequencies is small enough compared to the spectral range of interest, we implicitly assume that this first problem is negligible and that we can still use the identity theorem.
    \item Although the definition of an error function, such as the relative $L_2$ error, is possible, it is numerically impossible to construct a singularity expanded function with an infinite number of poles. This quickly leads to a truncation such as the one considered in Equation~\eqref{eq:5_3} for the SZF rather than the SEM expression. Consequently, the error cannot be reduced to $0$, and the unicity cannot be guaranteed. Adding more singularities does not result in a convergence towards the actual function, even though a better approximation in terms of error reduction.
    \item Measurements or simulations are always associated with errors, even though they can be extremely small compared to the actual value in the case of simulations. The result is that the poles (and zeros or residues) that we attempt to retrieve are those of a close approximation of the actual function rather than this function $h(\omega)$ itself.
\end{enumerate}

\begin{figure}[H]
    \centering
    \includegraphics[width=.98\textwidth]{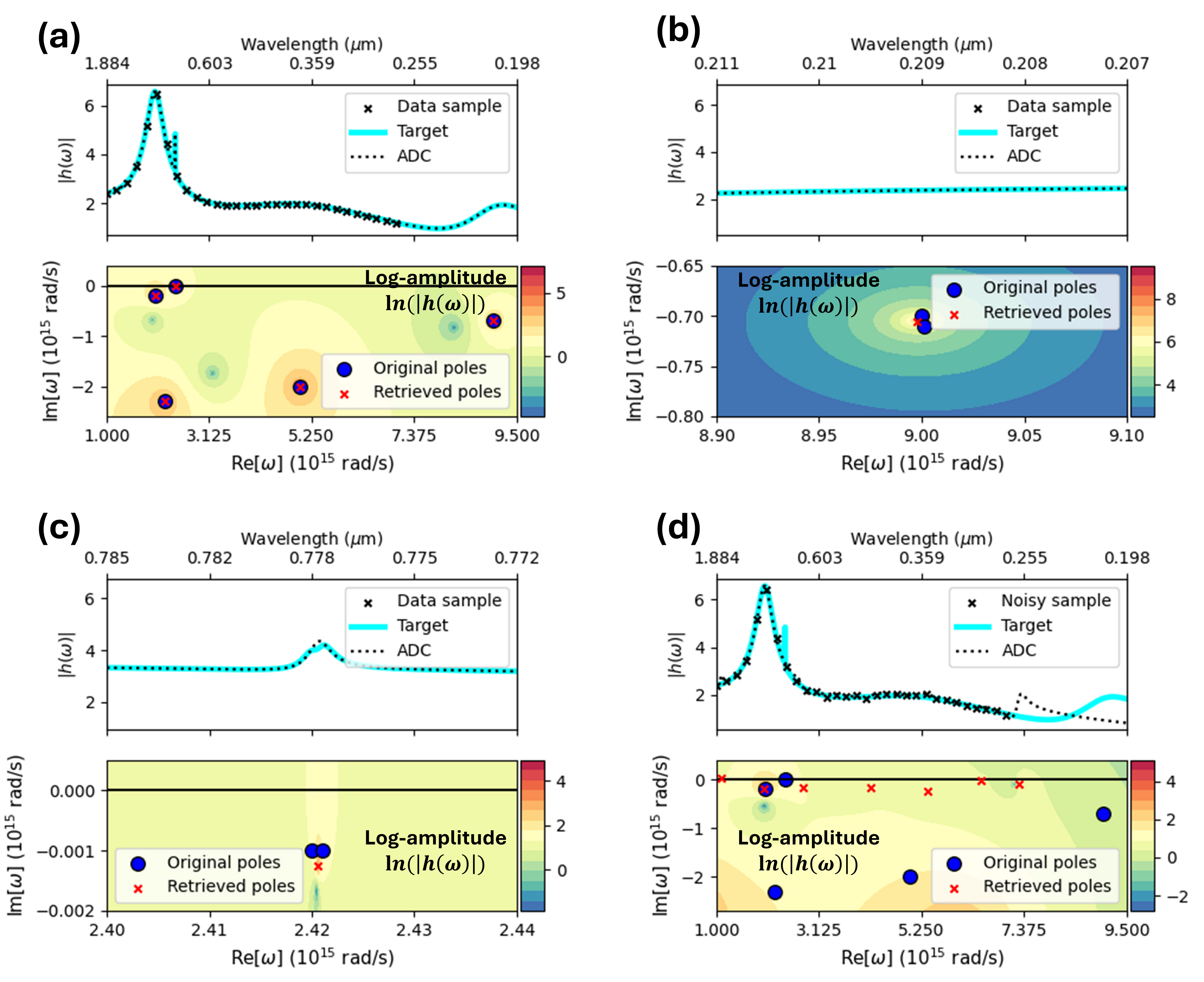}
    \caption{(a) Reconstruction of the test function $h(\omega)$ (cyan curve) described in Table~\ref{tab:5_1} using the ADC method (black dotted curve) on a sample of $30$ points (black crosses), without imposing the Hermitian symmetry, with a stability constraint coefficient $q_0=0.05$, with a maximum of $20$ poles, and with a maximum difference of $4$ between the numbers of poles and zeros. The poles of the target function (Original poles, blue) and those retrieved by the Cauchy method (Retrieved poles, red) are shown in the log-amplitude map, in the complex $\omega$ plane. (b) A pole $p=(9.001-.71\rmi)  \times 10^{15}$ rad/s is added at high frequencies, close to an original pole of $h(\omega)$. This results in an averaged retrieved pole rather than the retrieval of both poles. (c) A pole $p=(2.421-.001\rmi)  \times 10^{15}$ rad/s is this time added at low frequencies, resulting in a similar effect, \ie one average retrieved poles for two original poles. (d) A Gaussian noise is added to the sampled data, maintaining a Signal-to-Noise Ratio (SNR) of $30$. Only one pole, associated with the peak at around $2 \times 10^{15}$rad/s, is accurately retrieved.}
    \label{fig:5_9}
\end{figure}

\noindent
As we attempt to retrieve the poles and zeros or poles and residues, we are only able to recover an approximation of the function over a finite spectral window, with no guarantee that the obtained parameters offer a valid expansion of the data into the complex frequency plane. This is illustrated in Figure~\ref{fig:5_9}, where we apply the ADC method in four different situations on a test function $h(\omega)$ defined as
\begin{equation}
    h(\omega) = \sum_{\ell=1}^5 \frac{r_\ell}{\omega - p_\ell},
\label{eq:5_50}
\end{equation}
with the residues $r_\ell$ and poles $p_\ell$ given in Table~\ref{tab:5_1}. 

\begin{table}[h!]
    \centering
    \begin{tabular}{ | m{0.12\textwidth}<{\centering} | m{0.35\textwidth}<{\centering} | m{0.35\textwidth}<{\centering} | } 
        \hline
        \multicolumn{3}{|c|}{Test function $h(\omega)$} \\

        \hline
        Index $\ell$ & Pole ($\times 10^{15}$rad/s) & Residue ($\times 10^{15}$rad/s)\\
        
        \hline
        $1$ & $2-2\rmi$ & $e^{-\frac{1}{9}\rmi\pi}$\\
        
        \hline
        $2$ & $2.2 - 2.3\rmi$ & $e^{\frac{1}{9}\rmi\pi}$\\
        
        \hline
        $3$ & $2.42 - 0.002\rmi$ & $e^{\frac{17}{180}\rmi\pi}$\\
        
        \hline
        $4$ & $5 - 2\rmi$ & $e^{\frac{1}{9}\rmi\pi}$\\
        
        \hline
        $5$ & $9 - 0.7\rmi$ & $e^{\frac{1}{6}\rmi\pi}$\\
        
        \hline
    \end{tabular}
    % \captionsetup{justification=centering}
    \caption{Complex poles and residues of the test function $h(\omega)$ used in Equation~\eqref{eq:5_50}.}
    \label{tab:5_1}
\end{table}

In Figure~\ref{fig:5_9} (a), we consider the exact function and show that the target poles and residues are retrieved. In Figure~\ref{fig:5_9} (b) we add a pole close to the one outside of the sampled data, making it difficult to discriminate between the two. This results in an averaging of the poles (but not the residues). In Figure~\ref{fig:5_9} (c), a similar effect is obtained when adding a pole close to the one with the smallest imaginary part to $h(\omega)$.  In Figure~\ref{fig:5_9} (d) we randomly add noise to the data, with a noise-to-signal ratio of less than $.1\%$ for each value $h(\omega_i)$, resulting in a combination of the two previous effects.\\

\noindent
The actual poles hold important and rich information regarding a physical signal since they strongly contribute to its resonances~\cite{soltane2024extracting}. It is therefore necessary to be able to discriminate, amongst all the retrieved poles, between the natural poles which would be obtained if the identity theorem could hold, \ie with infinite precision over a continuous set of frequencies, and the effective poles, which result from the introduction of errors and only help in reducing the error between the truncated singularity expansion or SZF and the measured or simulated data.\\

\noindent
Under these considerations, we evaluate the efficiency of algorithms through four scores:
\begin{itemize}
    \item The overall efficiency estimated \via the relative $L_2$ precision $p_r^{(2)}=1-e^{(2)}$.
    \item The physical relevancy of the solutions through the ratio of Hermitian-symmetric to total poles $\rho_{\mathrm{herm}}=\frac{\mathrm{Hermitian~poles}}{\mathrm{retrieved~poles}}$.
    \item The stability of the solution with the ratio of stable to total poles $\rho_{\mathrm{stab}}=\frac{\mathrm{stable~poles}}{\mathrm{retrieved~poles}}$.
    \item The ratio of natural to total poles $\rho_{\mathrm{nat}}=\frac{\mathrm{ natural~poles}}{\mathrm{retrieved~poles}}$.
\end{itemize}
The computation of the first three scores is straightforward given a target curve and target poles. Nonetheless, the identification of natural poles is a complex problem. We calculate $\rho_{\mathrm{nat}}$ in two steps. Given a set of target natural poles $\mathcal{P}=\{p_1, ..., p_{M_p}\}$ and a set of retrieved poles $\mathcal{Q}=\{q_1, ..., q_{M_q}\}$, we calculate a relative distance matrix $\mathbf{D}=[D_{ij}]$ with $1 \leq i \leq M_p$ and $1 \leq j \leq M_q$, defined as
\begin{equation}
    D_{ij} = \min\left( \frac{|p_i - q_j|}{|p_i|}, \frac{|p_i - q_j|}{|q_i|} \right)
\label{eq:5_51}
\end{equation}
Using $\mathbf{D}$, we identify all the pairs of poles $(p_i, q_j)$ separated by a relative distance $\delta_{ij}$ shorter than an arbitrarily chosen threshold $\delta_0=10\%$. If $p_i$ and $q_j$ are close enough, we then decide on whether they are identical based on similarity of the resonances they would bring about in a response function if they were isolated. To do so, we first calculate the quality functions $\eta_i$ of $p_i$ and $\eta_j$ of $q_j$, which can be interpreted here as similar as the quality factors of the isolated resonances, but calculated over a spectral range~\cite{soltane2024extracting}:
\begin{gather}
    \eta_i(\omega) = \frac{\rmi}{4}\left[ \frac{\omega}{\omega-p_i} - \frac{\omega}{\omega-\overline{p_i}} \right] \\
    \eta_j(\omega) = \frac{\rmi}{4}\left[ \frac{\omega}{\omega-q_j} - \frac{\omega}{\omega-\overline{q_j}} \right]
\end{gather}
We then compute the standard deviation $\sigma_{ij}$ of the absolute difference of the quality functions evaluated at the $N$ sample frequencies $\omega_n$:
\begin{equation}
    \sigma_{ij} = \mathrm{st.dev.} \left( \{ \left|\eta_i(\omega_n)-\eta_j(\omega_n)\right| \}_{1\leq n \leq N} \right)
\end{equation}
The value of $\sigma_{ij}$ can be understood as a quantification of the maximum influence of the difference of the resonances brought about by the isolated poles $p_i$ and $q_j$. If $\sigma_{ij} < 2$, \ie if the difference of the resonances has a quality factor smaller than $2$, then we assume that $p_i$ and $q_j$ are identical, in which case the retrieved pole $q_j$ is counted as a natural pole if $p_i$ is a known target pole.

\subsection{Algorithm benchmarking}

\noindent
Using these criteria, we evaluate the performances of the ADC method and the auto-differentiation (AutoDiff) approach of SEMPO, which we compare to the original Cauchy method~\cite{sarkar_cauchy_2021}, as well as of two state-of-the-art methods, \ie the AAA algorithm~\cite{valera-rivera_aaa_2021,hofreither2021algorithm} and the Vector Fitting (VF) approach~\cite{gustavsen_rational_1999,gustavsen2006improving,deschrijver2008macromodeling}. The test function we want to fit is the Hermitian-symmetric version of function introduced in Equation~\eqref{eq:5_50}: 
\begin{equation}
    h(\omega) = \sum_{\ell=1}^5 \left[ \frac{r_\ell}{\omega - p_\ell} - \frac{\overline{r_\ell}}{\omega + \overline{p_\ell}} \right],
\label{eq:5_51_bis}
\end{equation}
where the residues $r_\ell$ and poles $p_\ell$ are those given in Table~\ref{tab:5_1}. We uniformly sample $h(\omega)$ to obtain a fitting dataset of $35$ frequencies $\omega_i$ ranging from $7\times10^{15}$rad/s to $1\times10^{15}$rad/s and their associated values $h_i=h(\omega_i)$. While this sample is used to run the algorithms, the relative $L_2$ precision is then calculated for a sample of $100$ points in the same spectral window. We test the performances of the different algorithms on noisy data $\tilde{h}_i$, obtained by adding Gaussian noise to $h_i$, with an SNR ranging from $10\log_{10}(50)$ to $10\log_{10}(1000)$.  The noisy data read as
\begin{equation}
    \tilde{h}_i = h_i + \sigma(b_{R,i} + \rmi b_{I,i}),
\label{eq:5_52}
\end{equation}
where
\begin{equation}
    b_{R,i},b_{I,i} \sim \mathcal{N}(0, 1).
\label{eq:5_53}
\end{equation}
We write the SNR as
\begin{equation}
    SNR = 10 \log_{10}\left(\frac{\displaystyle\sum_i|{h_i}|^2}{\displaystyle\sigma^2\sum_i|{b_{R,i}+\rmi b_{I,i}}|^2}\right)
\label{eq:5_54}
\end{equation}
allowing us to set the value of $\sigma$ \via
\begin{equation}
    \sigma = 10^{-SNR/20}\sqrt{\frac{\sum_i|{h_i}|^2}{\sum_i|{b_{R,i}+\rmi b_{I,i}}|^2}}. 
\label{eq:5_55}
\end{equation}
The SNR is set successively to $10\log_{10}(50)$, $10\log_{10}(100)$ and $10\log_{10}(1000)$ during the benchmarking.\\

\noindent
We perform a parameter sweep for all the aforementioned pole retrieving methods. Each set of parameter is tested with different levels of Gaussian noise in the test data as the SNR is increased from $10\log_{10}(50)$ to $10\log_{10}(1000)$. The VF approach works by fitting the data with an expression similar to the truncated singularity expansion in Equation~\eqref{eq:5_1_sem}, to which is optionally added a term proportional to the frequency $\omega$. We run the algorithm without this additional term, and only sweep through the number of complex Hermitian-symmetric poles and the number of purely imaginary poles:
\begin{itemize}
    \item The number of complex Hermitian-symmetric poles ranges from $4$ to $18$ with a step of $2$.
    \item The number of purely imaginary poles ranges from $0$ to $4$ with a step of $1$.
\end{itemize}

\noindent
The AAA algorithm fits the data with a stable rational approximation that takes the form of a ratio of partial fractions. This rational approximation can be recast as a truncated SZF, allowing the AAA algorithm to provide the poles and zeros of the fitted function. Only one parameter corresponding to the maximum number of terms in the partial fractions is swept through:
\begin{itemize}
    \item The maximum number of terms ranges from $2$ to $14$ with a step of $1$.
\end{itemize}

\noindent
In our implementation of the original Cauchy method described in Section 2.1, only the maximum number of poles is swept through:
\begin{itemize}
    \item The maximum number of poles ranges from $8$ to $20$ with a step of $2$.
\end{itemize}

\noindent
We run the ADC method with an imposed Hermitian symmetry, and a maximum difference of $4$ between the numbers of poles and zeros. We sweep through the maximum number of poles and the stability constraint coefficient $q_0$:
\begin{itemize}
    \item The maximum number of poles ranges from $8$ to $20$ with a step of $2$.
    \item The stability coefficient $q_0$ takes $10$ evenly spaced values from $0$ to $0.05$.
\end{itemize}

\noindent
For the auto-differentiation approach, the hyperparameters are fixed: the learning rate is set to $.007$, and the number of iterations to $22000$, which is about two third of the value classically used in the optical structures of the previous structures. The swept-through parameters are the number of complex Hermitian-symmetric poles and the number of purely imaginary poles:
\begin{itemize}
    \item The number of complex Hermitian-symmetric poles ranges from $2$ to $10$ with a step of $1$.
    \item The number of purely imaginary poles goes from $0$ to $2$ with a step of $1$.
\end{itemize}

\noindent
Let us point out that we do not consider here the combined chained-ADC and auto-differentiation approach as it is primarily used in wide spectral range exhibiting rapid variations. The chained ADC method presents no advantage with the test function used for the benchmarking as it introduces several singularities which cannot be all removed at the seams of the sub-windows, despite the merging routine described in Section 4.\\

\noindent
Other methods operating in the temporal domain such as the harmonic inversion~\cite{mandelshtam_harmonic_1997} or the Prony-SS decomposition method~\cite{cong2022frf} are not tested here, since they involve a Fourier transform to obtain temporal data, which might lead to additional sources of error difficult to account for. Nonetheless, these temporal approaches work by fitting time-decaying exponentials corresponding to the analytical inverse Fourier transform of the singularity expansion. We thus expect them to perform similarly to SEMPO. In addition, the benchmarking scripts are available on the Zenodo repository to allow for additional performance 
evaluations with other test functions or algorithms.\\

\begin{figure}[H]
    \centering
    \includegraphics[width=.98\textwidth]{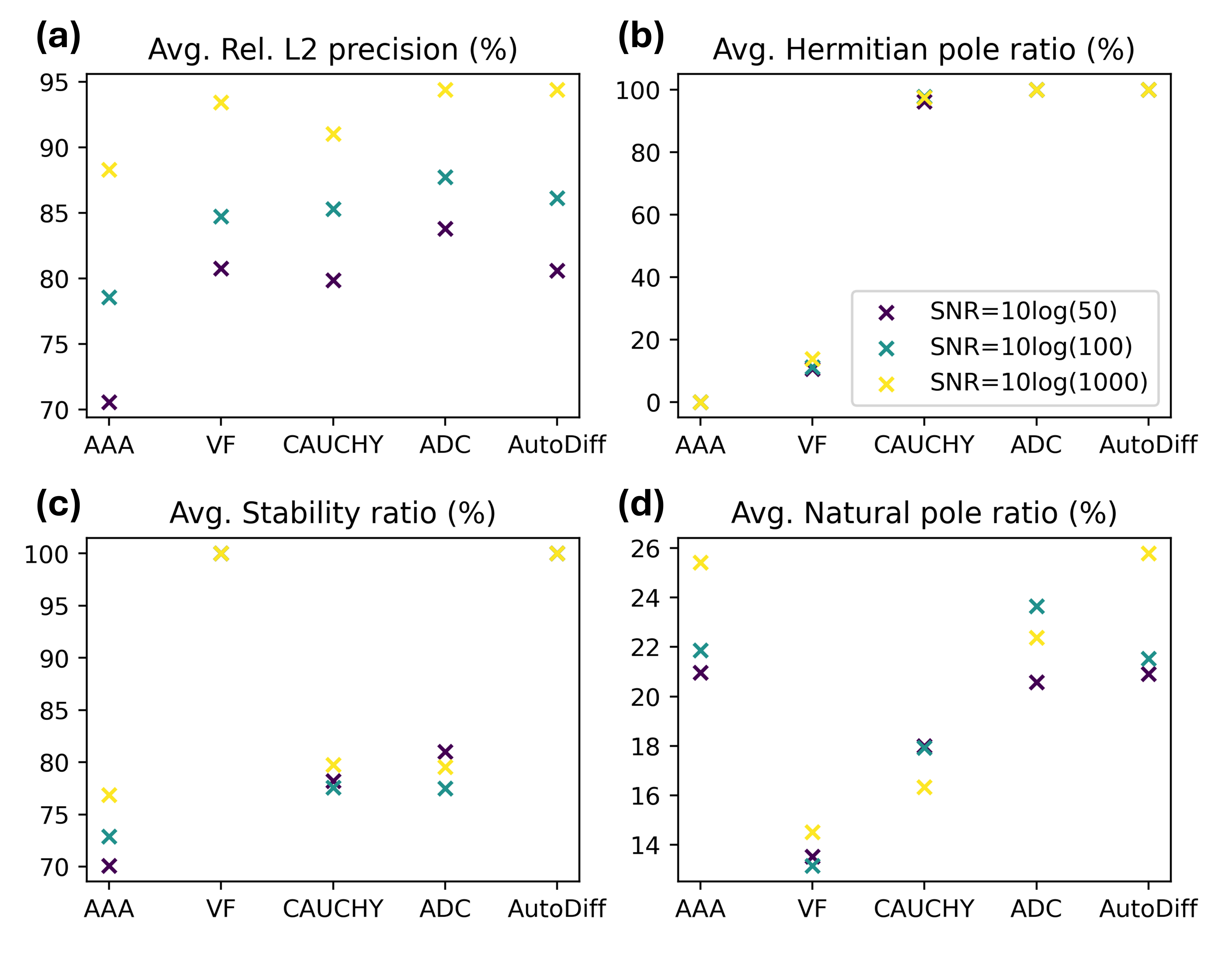}
    \caption{Benchmarking results of the AAA algorithm, VF algorithm, original Cauchy method (CAUCHY), ADC method and auto-differentiation (AutoDiff) approach, for an SNR of $10\log_{10}(50)$, $10\log_{10}(100)$ or $10\log_{10}(1000)$. (a) Average relative $L_2$ precision. (b) Average Hermitian-symmetric pole ratio. (c) Average stable pole ratio. (d) Average natural pole ratio. The averaging is performed over the set of parameters for each SNR level.}
    \label{fig:5_10}
\end{figure}

\noindent
As previously mentioned, we filter out all the algorithm runs performed with an SNR $\leq10\log_{10}(10)$, and calculate the four scores defined in the previous Subsection 5.1 for each of the remaining runs. The results are presented in Figure~\ref{fig:5_10} where they are grouped up based on the SNR and the algorithm or method used. The average relative $L_2$ precision $p_r^{(2)}$ is in Figure~\ref{fig:5_10} (a), the Hermitian ratio $\rho_{\mathrm{Herm}}$ in Figure~\ref{fig:5_10} (b), the stability ratio $\rho_{\mathrm{stab}}$ in Figure~\ref{fig:5_10} (c), and the Natural pole ratio $\rho_{\mathrm{nat}}$ in Figure~\ref{fig:5_10} (d). The averaging is performed over the set of parameters for each SNR level. For a given SNR, we generate $50$ datasets by drawing $50$ times the noise. The scores for a given algorithm, with a given set of parameters is thus an average over these $50$ drawings. Let us point out that the stability ratio and Hermitian ratio are biased towards SEMPO since it has been developped in order to account for these physical constraints. Their presence is however helpful in understanding the advantage of some methods compared to others.\\

\noindent
In terms of average $L_2$ precision, we find that the VF algorithm, the ADC method and the AutoDiff approach perform similarly, as observed in Figure~\ref{fig:5_10} (a). The ADC ranks first at SNR levels below $10\log_{10}(1000)$, but is equivalent to the AutoDiff and VF as the SNR reaches $10\log_{10}(1000)$. This result is in agreement with the design of the ADC method, which builds upon the classical Cauchy method while minimizing the relative error. Let us point out that by increasing the value of the hyperparameter corresponding to the number of iterations from $20000$ (used for the benchmarking) to $35000$ (usual parameter), the average precision of the AutoDiff can be increased at all SNR. This shows the sensitivity of the AutoDiff approach to the hyperparameters.\\

\noindent
The high precisions of the VF algorithm and ADC methods comes at the expense of non-physical solutions. The VF algorithm does not comply with the Hermitian-symmetry as shown in Figure~\ref{fig:5_10} (b), while the stability of the solutions is not fully ensured in the ADC method as seen in Figure~\ref{fig:5_10} (c). Only the AutoDiff approach yields solutions fully respecting both criteria.\\

\noindent
Finally, we observe that the AutoDiff approach offers a more robust way of retrieving the actual poles of a system as indicated by the higher natural pole ratio in Figure~\ref{fig:5_10} (d) with respect to other methods, at all SNR. This ratio is obtained by dividing the number of natural poles by the number of retrieved poles instead of actual poles. This allows us to reward methods which provide fewer poles, making it more likely to identify them when the target is unknown, \ie in classical applications.\\

\noindent
Overall, the ADC method provides a good alternative to the VF algorithm for obtaining a Hermitian-symmetric solution with high precision, but it lags behind in stability. On the other hand, the AutoDiff approach provides both accurate solutions and Hermitian-symmetric, stable poles, while showing the best performance in the retrieval of natural poles of the system bringing about the signal.

% A RETIRER =================================================================
% \noindent
% The identification of natural poles in the practical and common cases where no analytical expressions of $h(\omega)$ are known rises another complex problem which has yet to solved. Inspired by the work performed in [REF], we define natural poles as the set of overlapping or common poles $\mathcal{P}_{c}$ between the sets of poles $\mathcal{P}_i,~1\leq i\leq N_A$, obtained \via $N_A$ pole retrieving methods. The higher the number $N_A$ of algorithms is, the more likely we are to avoid identifying effective poles as natural poles, provided that efficient algorithms are selected. Nonetheless, some natural poles are generally omitted, and there is no guarantee that the identified natural poles are in fact natural poles, as previously stated. 
% ===========================================================================

\newpage

\section{Conclusion}

\noindent
To conclude, we described the two approaches used in the SEMPO toolbox to retrieve complex poles, zeros and residues of arbitrary response functions. By considering the response functions of different optical systems, we showed that the improved version of the Cauchy method and the auto-differentiation based approach allow for highly accurate reconstructions of the response functions by retrieving the complex parameters of the SEM pole expansion and the SZF.\\

\noindent
We explained how the Accuracy-Driven Cauchy method (ADC method) improves the accuracy of the reconstructions and can be constrained to account for the stability and Hermitian-symmetry of physical systems.\\

\noindent
We introduced an equivalent form of the SEM, namely the GDL model, which offers a different interpretation of response functions, and can still be retrieved using the auto-differentiation-based approach. This method relies on a weighted sum of loss function terms which can be used to give more importance to the average reconstruction of a target function, the maximum deviation from the target, the real part of the response function, or the imaginary part.\\

\noindent
Although the ADC method can run much faster than the auto-differentiation approach, it is not as efficient when strong variations are observed in the response curve. As for the auto-differentiation method, it is highly dependent on the initial distributions of poles and residues which can slow down the optimization process. Consequently, we have shown how to associate the two approaches sequentially to improve the accuracy of the reconstruction, speed up the optimization process, and reduce the size of the distribution of poles.\\

\noindent
\revision{We provide a thorough comparison between the performances brought by the ADC method and the auto-differentiation approach of SEMPO to other state of the art algorithms. Results and analysis show that while ADC and auto-differentiation feature comparable performances with other methods in terms of relative $L_2$ error, they outperform other methods in providing stable and Hermitian-symmetric solutions as well as in retrieving the natural poles of a system.}\\

\noindent
While we have considered optical responses of photonic nanostructures to study and assess the performances of SEMPO, this method is general and can be applied to other fields of physics since it applies for any spectral response or transfer function in wave physics.

\section*{Code availability}
SEMPO is available in the Zenodo repository in reference \cite{bensoltane_zenodo_code}

\bibliographystyle{unsrt}
\bibliography{biblio}

\end{document}

\begin{frontmatter}

\title{Supporting Information for the manuscript ``SEMPO - Retrieving complex poles, residues and zeros from arbitrary real spectral responses''}

\author[1]{Isam BEN SOLTANE\corref{cor1}}
\ead{isam.ben-soltane@fresnel.fr}
\author[1]{Mahé ROY}
\author[1]{Rémi ANDRE}
\author[1]{Nicolas BONOD\corref{cor1}}
\ead{nicolas.bonod@fresnel.fr}

\affiliation[1]{Aix Marseille Univ, CNRS, Centrale Mediterranee, Institut Fresnel, 13013 Marseille, France}

\cortext[cor1]{Corresponding authors}

\end{frontmatter}

\section{Benchmarking of the algorithms}

\begin{figure}[H]
    \centering
    \includegraphics[width=.8\textwidth]{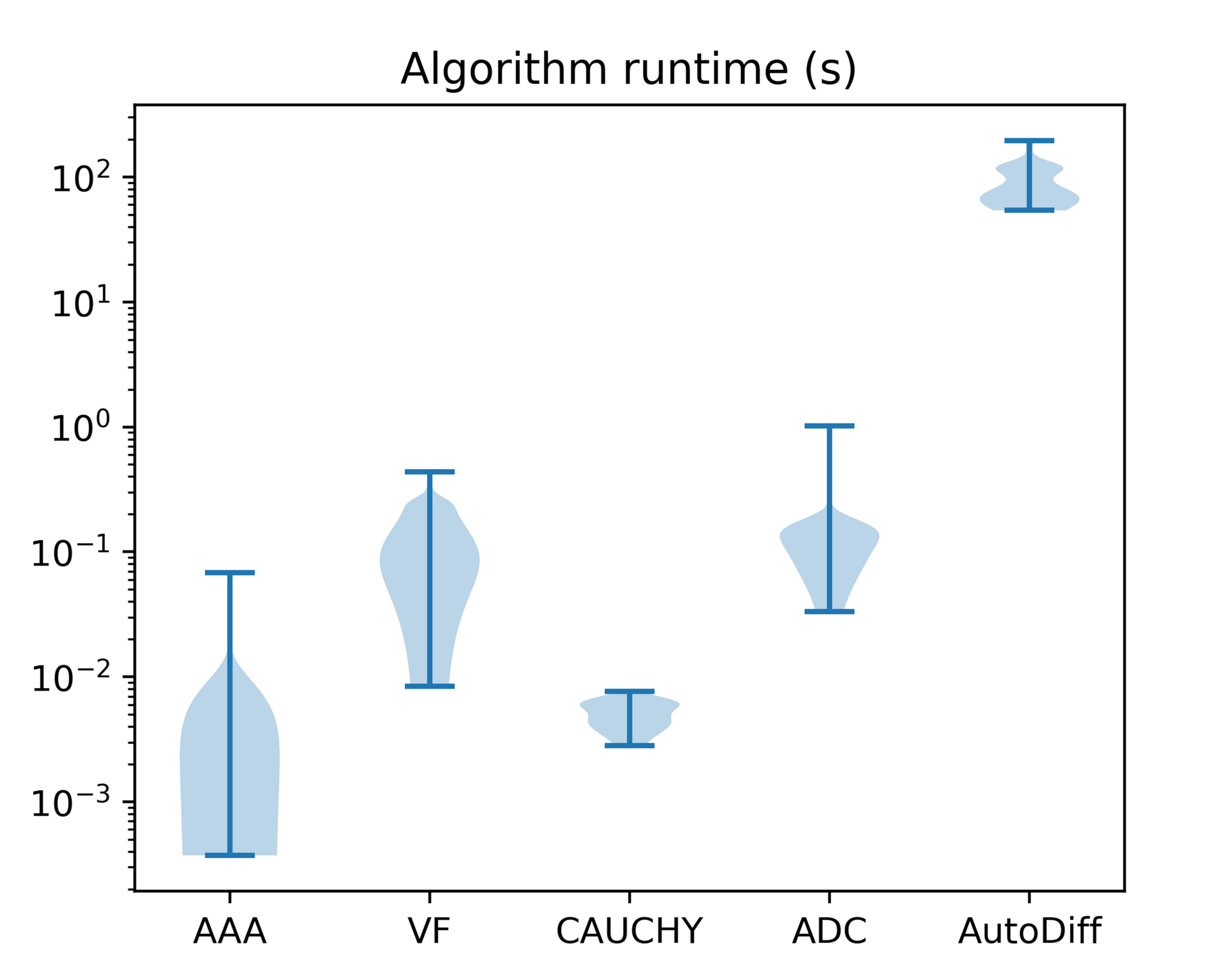}
    \caption{Runtime of the algorithms benchmarked in the manuscript in seconds.}
    \label{fig:5_S1}
\end{figure}

\noindent
Section 5 of the manuscript is dedicated to the evaluation of the performance of the Accuracy-Driven Cauchy (ADC) method and the Auto-Differentiation (AD) approach from SEMPO, as well as the Adaptive Antoulas-Anderson (AAA) method, the Vector Fitting (VF) method, and our implementation of the original Cauchy method. These different methods were tested with different sets of parameters and noise levels. In addition to the metrics presented in the manuscript, we provide an overview of the runtime associated with the different runs considered in Figure~\ref{fig:5_S1}. Let us point out that the algorithms ran on a sample of size $35$.

\noindent
While the AAA method systematically runs in less than $10$ms, the VF, original Cauchy and ADC methods can reach a runtime of $1$s or more. We run the AutoDiff method with at least $30 000$ iterations to ensure that the cost function remains at a stable low level. Under this configuration, each run lasts at least $45$s. Combining the chained ADC approach with the AutoDiff, as described in section 4 of the main manuscript, often allows for starting with a better initialisation which reduces the number of iterations by a factor of two.

\section{SEMPO time scaling}

\begin{figure}[H]
    \centering
    \includegraphics[width=.98\textwidth]{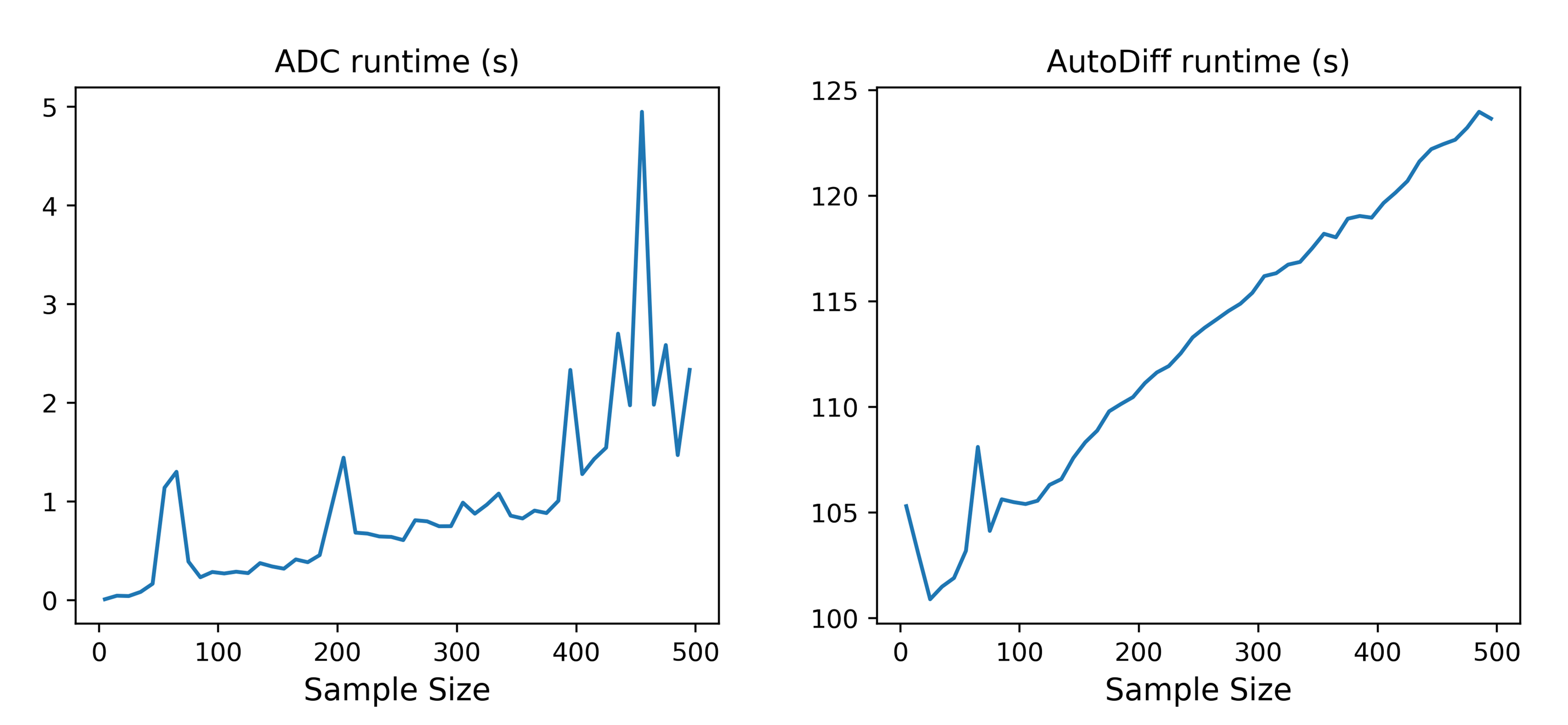}
    \caption{Runtimes of the ADC and AutoDiff methods when the sample size is increased. The parameters of the ADC method are set to $20$ maximum poles, a maximum difference between the numbers of poles and zeros of $2$, and a stability coefficient $q_0=0.005$. The AutoDiff is applied with $8$ Hermitian-symmetric complex poles and $3$ purely imaginary poles.}
    \label{fig:5_S2}
\end{figure}

\noindent
The datasets used in the manuscript are systematically sampled to fewer than $100$ points. We present in Figure~\ref{fig:5_S2} an overview of the time scaling of the ADC and AutoDiff approach when the sample size is increased. The parameters of the algorithms themselves are fixed during the runs. We use $20$ maximum poles, a maximum difference between the numbers of poles and zeros of $2$, and a stability coefficient $q_0=0.005$ for the ADC method. For the the AutoDiff approach, $8$ Hermitian-symmetric complex poles and $3$ purely imaginary poles are used.

\section{Information regarding the computer used}

All calculations were performed on a laptop with an AMD Ryzen 5 PRO 8540U, and 32GB of SODIMM DDR5 RAM.